\documentclass[10pt,prc,aps]{revtex4}
\draft

\usepackage{amsmath}
\usepackage{graphicx}
\usepackage{subfigure}
\usepackage{color}
\usepackage{mathtools}
\begin{document}

\title{ Tidal deformability in neutron stars from a microscopic point of view}  
\author{            
Francesca Sammarruca\footnote{Corresponding author. Email: fsammarr@uidaho.edu} and Prabin Thapa}                                                         
\affiliation{ Physics Department, University of Idaho, Moscow, ID 83844-0903, U.S.A. 
}

\date{\today} 

\begin{abstract}
We  present results for the tidal deformability in neutron stars, the tidal Love number $k_2$, and the effective deformability of a binary system. The microscopic equation of state for cold $\beta$-stable neutron matter is based upon high-precision two-neutron forces and includes the chiral three-neutron forces required at the chosen order. We review and motivate our choices for the high-density continuation of the microscopic equation of state. We discuss our predictions and observe that they are well within multimessenger constraints. In contrast, stiff equations of state that yield radii larger than about 13 km are ruled out by GW170817 constraints.

\noindent 
{\bf Keywords:} Tidal deformability; neutron matter; neutron stars; chiral effective field theory
\end{abstract}
\maketitle

\section{Introduction} 
\label{Intro} 

A fully microscopic equation of state (EoS) up to central densities of the most massive stars -- potentially
involving non-nucleonic degrees of freedom and phase transitions -- is not within reach.
Nevertheless, neutron stars are powerful natural laboratories for constraining theories of
the nuclear EoS~\cite{Abb17a, Abb17b, Abb18, Abb19, Mil19, Mil21} and studying matter under extreme conditions.

A mix of all fundamental forces existing in nature is present in a neutron star, which continues to motivate the large interdisciplinary interest in its
properties and the intense effort to constrain those properties both
observationally and experimentally, with various levels of statistical and systematic uncertainties, see e.g. Ref.~\cite{Soum+18}.
When working with microscopic predictions (as opposed to phenomenological models or statistical analyses), one 
 must be mindful of the theory's limitations and identify the best ways to extract and interpret
 information from the observational or terrestrial data.

With the first detection of gravitational waves (GW) on
September 14, 2015 from LIGO, major accomplishments
have been made within the fields of physics and
astronomy~\cite{MY2019}. Gravitational waves tied up one of the
last loose ends in the theory of general relativity with
remarkably high precision~\cite{Cerv+2016}. Our present understanding of black holes and neutron stars
also rely heavily on data collected from GW. 

Just as black holes release immense gravitational
waves, neutron stars are remarkable sources
of these waves. Exotic processes are triggered within a
neutron star's core due to the enormous pressure,
which can forcefully eject mass at relativistic speeds.
Consequently, gravitational waves become vital means
of measuring these fascinating astrophysical objects and
the dynamic processes within them~\cite{Lasky2015}. 
For gravitational waves to produce discernible signals, 
 a crucial requirement is the presence of
asymmetrical deformations in the neutron star. 
Alternative mechanisms, such as pulsar “glitches," 
events where neutron stars undergo a sudden “jump”
in spin frequency, or magnetar flares, where a neutron
star with an exceptionally high magnetic dipole experiences
magnetic decay, can also induce substantial gravitational
waves~\cite{Lasky2015}.

During a neutron star merger event, the gravitational field of one neutron star on
the other induces tidal deformations, which are encoded in the late-inspiral gravitational
wave signal. More precisely, when a neutron star is placed in the perturbing tidal gravitational
field generated by its companion compact star, its shape becomes distorted, resulting in an induced quadrupole moment.
The induced
quadrupole moment of the neutron star will impact the energy
of the system and increase the rate of emission of gravitational waves, with the
 result that the binary system emits more energy and evolves faster towards
the collision and merger~\cite{FH2008,Hind2008}. Clearly, compact binaries that are detectable through gravitational waves are natural systems to observe this  type of interaction, which opens the way to finding additional constraints to the neutron star EoS.  A stiffer EoS, where the pressure increases rapidly with 
density, yields stars with larger radii and larger tidal effects. Therefore, tidal distortion reflects 
the stiffness (softness) of the EoSs, with the inspiral proceeding more rapidly for stars with
larger tidal deformability.

In this paper, we focus primarily on the tidal properties
of neutron stars, which have a direct impact on the emitted gravitational wave
signal, offering the opportunity of detecting the latter to constrain the former~\cite{FR2012,BR2017,Bai2019}. This prospect was first realized in August of 2017, with the detection of gravitational waves generated by a coalescing neutron star binary. This groundbreaking event, known as GW170817~\cite{Abb17a}, 
 was made possible 
by the second-generation ground-based detectors LIGO~\cite{AL} and Virgo~\cite{Acern+2015}.  We can expect future observations to detect the inspiral and
coalescence of more compact binary systems. 

As we review below, for a given neutron star mass, the tidal deformability
is strongly correlated with the star radius and, thus, the star compactness, C, defined as M/R. Keeping in mind that accurate and precise neutron star radius measurements are challenging~\cite{Khad+2025}, the importance of tight constraints on global neutron star properties that depend on M/R is self-evident. We also recall that the radius of the canonical neutron star correlates with the (model dependent) slope of the symmetry energy around saturation density.

\begin{figure*}[!t] 
\centering
\hspace*{1cm}
\includegraphics[width=8.0cm]{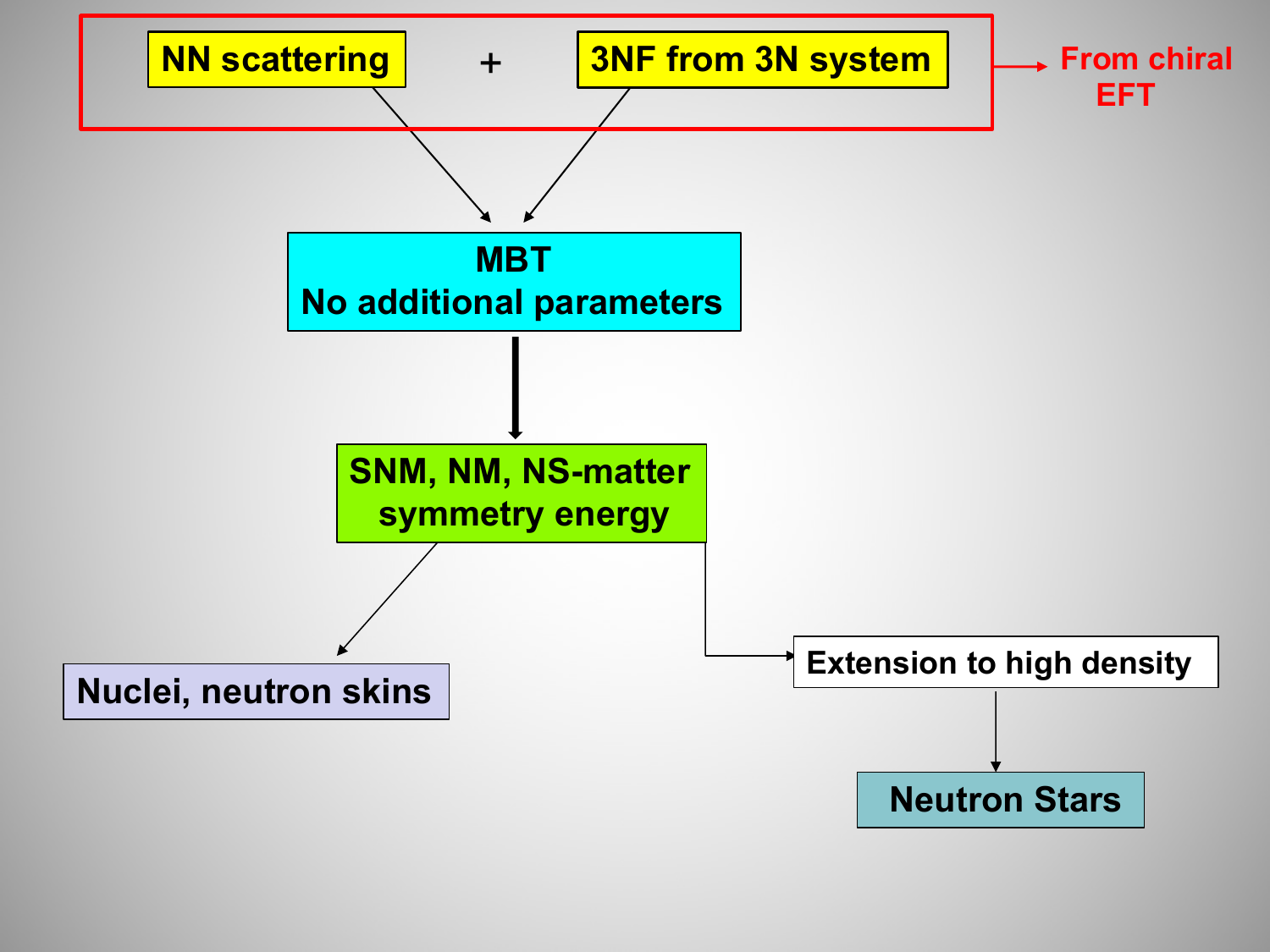}\hspace{0.01in} 
\vspace*{0.2cm}
\caption{(Color online) Schematic representation of our {\it ab initio} pipeline. 
}
\label{path}
\end{figure*}

Although this paper could be seen as an extension of Ref.~\cite{S+A_2025}, we note that it is an important extension. The tidal deformability encodes the impact of the EoS on the phase evolution of the waveform, which we have not previously confronted with our EoS. Tidal deformation is yet another astrophysically observable macroscopic property of neutron stars that can be used to study their interiors. Currently, Bayesian statistical analyses are very popular in the nuclear physics community, as they should be. Our approach is not a statistical analysis, in that we make predictions remaining as close as possible to the chiral theory of nuclear forces, and follow new constraints as they become available. We believe that our findings in Ref.~\cite{S+A_2025} are valuable and should be tested against these independent data.
Furthermore, while this paper focuses on a particular, fundamentally important physical property of neutron stars, we wish to place it  within our ongoing microscopic studies of the nuclear equation of state, where we keep a consistent path from free-space nucleon-nucleon ($NN$) data to neutron-rich nuclei and to neutron stars, as illustrated in Fig.~\ref{path}. ``MBT" stands for many-body theory, while SNM, NM, and NS signify symmetric nuclear matter, neutron matter, and neutron-star matter, respectively.

This paper is organized as follows. In Sec.~\ref{TF}, we review briefly our theoretical ingredients, omitting details that have been published elsewhere. In Sec.~\ref{high}, we construct high-density continuations of the microscopic EoS, paying particular attention to the speed of sound in stellar matter. In Sec.~\ref{gen}, we outline the calculation of the tidal deformability, which can be easily incorporated in the solution scheme used for the well-known Tolman-Oppenheimer-Volkoff (TOV) differential equations for $\frac{dP}{dr}$ and $\frac{dm}{dr}$. Our results, along with other predictions and constraints, are discussed in Sec.~\ref{res}. Our main takeaways are summarized in Sec.~\ref{concl}.

\section{Theoretical tools}
\subsection{Basic concepts in chiral effective field theory}
\label{TF} 
The first step in a development of an EFT is the identification of a ``soft scale'' and a ``hard scale.'' For this purpose, guidance can be found in the hadron spectrum, where a large separation exists between the mass of the pion and the mass of the vector meson $\rho$. It is therefore natural to identify the pion mass as the soft scale while the mass of the $\rho$ sets the hard scale, approximately 1 GeV. Moreover, since quarks and gluons are ineffective degrees of freedom in the low-energy regime, pions and nucleons are the appropriate degrees of freedom of the EFT. The connection between QCD and the EFT is established through the symmetries of low-energy QCD.  At this point, we can write the most general Lagrangian consistent with those symmetries (and  their breaking). Following the prescription of the theory as reviewed in Ref.~\cite{ME11}, the QCD Lagrangian is given by:

\begin{equation}
\mathcal{L} = \bar{q}(i\gamma^{\mu}\mathcal{D}_{\mu} - \mathcal{M})q - \frac{1}{4} \;  \mathcal{G}_{\mu\nu,a}\mathcal{G}^{\mu\nu}_{a} \; ,
\end{equation}
where `$q$' is the quark field, $\mathcal{D}_{\mu}$ represents the gauge covariant derivative, $\mathcal{M}$ is the quark mass matrix, and  $ \mathcal{G}^{\mu\nu}_{a}$ is the gluon strength field tensor.

Chiral symmetry is conservation of ``handedness," and is an exact symmetry for massless particles. Chiral symmetry occurs in the limit of vanishing quark masses, which amounts to dropping the quark mass matrix term in the above Lagrangian. In fact, such term is responsible for \textit{explicit} breaking of chiral symmetry, as can be seen from the following.
The quark mass matrix,
\begin{equation}
\mathcal{M} = 
\begin{pmatrix}
m_u & 0 \\
0 & m_d 
\end{pmatrix}
\end{equation}
can be recast in terms of the identity matrix and the third Pauli spin matrix:

\begin{equation}
\mathcal{M} = \frac{(m_{u} + m_{d})}{2}
\begin{pmatrix}
1 & 0 \\
0 & 1
\end{pmatrix}
+\frac{m_{u}-m_{d}}{2}
\begin{pmatrix}
1 & 0 \\
0 & -1
\end{pmatrix}
\; .
\end{equation}
Clearly, the first term respects isospin symmetry while the second term vanishes if the masses of the ``up" and ``down" quarks are equal. Thus, the small difference in the quark masses breaks isospin symmetry.
On the other hand, the expression above breaks chiral symmetry explicitly as a result of the non-zero quark masses. However, since the masses of the ``$u$" and ``$d$" quarks are very small compared to typical hadronic masses, explicit breaking of chiral symmetry is a small effect.

Next, we briefly address the {\it spontaneous} breaking of chiral symmetry, for which there is clear evidence in the hadron spectrum. The spontaneous breaking of a global  (as opposed to local) symmetry is accompanied by the appearance of a so-called  massless ``Goldstone Boson."The particle which fulfills these requirements is the pion, an isospin triplet pseudoscalar boson.
The pion is light relatively to the other mesons in the hadron spectrum but not massless, which is due to the explicit chiral symmetry breaking from the non-vanishing quark masses.

Having identified pions and nucleons as the appropriate degrees of freedom, one  can then proceed to construct the Lagrangian of the effective theory:

\begin{equation}
\label{lagr}
\mathcal{L}_{eff} = \mathcal{L}_{\pi\pi} + \mathcal{L}_{\pi N} + \mathcal{L}_{NN} + ...
\end{equation}

 This effective Lagrangian is expanded in the form of the ``soft scale'' over the ``hard scale'',  $\frac{Q}{\Lambda_{\chi}}$. $Q$ is of the order of the pion mass, whereas $\Lambda_{\chi}$ is the energy scale of chiral symmetry breaking, approximately 1 GeV.

 Through a scheme known as power counting, the most important contributions to the effective Lagrangian are accounted for first, with increasing order resulting in consistently smaller terms. While the expansion itself is, of course, infinite, at each order we are assured that the number of terms to be retained is finite and well defined. 

\subsection{Chiral truncation error}
\label{error}
 Crucial to chiral EFT is the trucation error, defined below
 If observable $X$ is known at order $n$ and at order $n+1$, a reasonable estimate of the truncation error at order $n$ can be expressed as the difference between the value at order $n$ and the one at the next order:
\begin{equation}
\Delta X_n = |X_{n+1} - X_n| \; ,
\label{del} 
\end{equation} 
since this is a measure for what has been neglected at order $n$.
To estimate the uncertainty at the highest order that we consider, we follow the prescription of Ref.~\cite{EKM15a}. For an observable $X$ that depends on the typical momentum of the system under consideration, $p$, one defines $Q$ as the largest between $\frac{p}{\Lambda_b}$ and $\frac{m_{\pi}}{\Lambda_b}$, where $\Lambda_b$ is the breakdown scale of the chiral EFT, for which we assume 600 MeV. The uncertainty of the value of $X$ at N$^3$LO is then given by:
\begin{displaymath}
\Delta X = \max \{Q^5|X_{LO}|, Q^3|X_{LO} - X_{NLO}|,Q^2|X_{NLO} - X_{N^2LO}|, 
\end{displaymath}
\begin{equation}
Q|X_{N^2LO} - X_{N^3LO}| \} \; ,
\label{err}
\end{equation} 
where $p$ could be identified with the Fermi momentum at the density under consideration.

We close this section with a warning.
While the predictions at N$^2$LO are fully {\it ab initio}, 
a warning is in place for current N$^3$LO calculations.
As pointed out in Ref.~\cite{EKR20}, 
there is a problem with the regularized three-nucleon-force (3NF) at N$^3$LO (and higher orders):
in all present nuclear structure calculations. 
The N$^3$LO 3NFs currently in use are regularized
by a multiplicative regulator applied to the 3NF expressions as
derived from dimensional regularization.
This approach leads to violation of chiral symmetry at N$^3$LO
and destroys the consistency between two- and three-nucleon forces~\cite{EKR20, Ep+22}. Consequently, none of the current calculations that
include 3NFs at N$^3$LO (and beyond) can be considered truly {\it ab initio}.
An appropriate symmetry-preserving regulator~\cite{EKR20} should be applied to  
 the 3NF at N$^3$LO from Refs.~\cite{Ber08,Ber11}. 
At the present time, reliable predictions exist only at 
N$^2$LO, NLO, and LO. However, for the few fully {\it ab initio} calculations, the precision at N$^2$LO is unsatisfactory. A first step towards deriving consistently regularized nuclear
interactions in chiral EFT has been proposed in Refs.~\cite{KE23a,KE23b}. It requires the cutoff to be introduced already at the level of the effective Lagrangian. A path integral approach~\cite{KE23a}
can then be applied to the regularized chiral Lagrangian to derive nuclear forces through 
the standard power counting of chiral EFT.

\section{The speed of sound in neutron star matter} 
\label{high}

\subsection{Subluminar and non-conformal}
\label{non_conf}

The content of this section is based on Ref.~\cite{S+A_2025}, where we 
discussed how causality and maximum-mass constraints pose considerable restrictions on the general features of the high-density EoS, if such continuation follows an ab intio portion. We also recall that our matching densities are $\rho_1$ = 0.277 $fm^3$ and
$\rho_2$ = 0.506 $fm^3$.

 Parametrizations of the high-density EoS in terms of the speed of sound can be used as an alternative to, or in combination with the popular polytropic extensions.
Defining $\rho_0$,  $\epsilon_0$, and $P_0$  as the baryon density, the energy density, and the pressure at threshold (the density at which the EoS parametrization is attached to the previous piece), we write~\cite{Kan+21, Tew18}, for $i \ge 1$,
\begin{equation}
\rho_i = \rho_{i-1} + \Delta \rho \; ,
\end{equation}
\begin{equation}
\epsilon_i = \epsilon_{i-1} + \Delta \epsilon     \; ,
\end{equation}
and 
\begin{equation}
 \Delta \epsilon= \Delta \rho \frac{\epsilon_{i-1} + P_{i-1}}{\rho_{i-1}}  \; .
\end{equation}

A reasonable parametrization for the speed of sound is:
\begin{equation}
\Big (\frac{v_s}{c} \Big )^2_i = 1 - c_1 exp \Big [ - \frac{(\rho_i - c_2)^2}{w^2}\Big ] \; , 
\label{vs}
\end{equation}
where the constants $c_1$ and $c_2$ are determined from continuity of the speed of sound and its derivative at the threshold density. We find this to be a practical choice, not the only one, for a function that should approach a limit asymptotically. The main point is to preserve causality, which could be done, for instance, by attaching additional polytropes with appropriate steepness so as to control maximum mass and causality.

 Using the relation between the dimensionless speed of sound squared and $dP/d\epsilon$, we write the pressure above the threshold as
\begin{equation}
P_i = \Big ( \frac{v_s}{c} \Big )^2_{i-1} \Delta \epsilon + P_{i-1} \; ,
\label{pi}
\end{equation}
This EoS continuation is obviously causal at all densities and allows for a maximum mass of 2.07 $M_{\odot}$.

We found~\cite{S+A_2025} that a convenient ansatz, for the purpose of achieving high maximum masses while respecting causality at any density, is to first continue the EoS with a relatively steep polytrope,
\begin{equation}
P(\rho) = \alpha \rho^{\gamma} \; ,
\end{equation}
 and then follow with a parametrization as in Eqs.~(\ref{vs}--\ref{pi}), which will remain causal by construction. The matching densities are 0.277 $fm^{-3}$ and 0.563 $fm^{-3}$~\cite{S+A_2025}.
We underlined that this scenario is fundamentally related to the softness of the chiral predictions. Therefore, the nature of the predictions at normal density have a far-reaching impact, which extends to densities up to a few times normal density.

That being said, piecewise polytropic extensions are quite general and remain a popular extension method. We found that, leveraging the flexibility offered by Eq.(12) over a limited high-density interval, would allow us to control the maximum mass, whereas the causality-preserving pressure-density relation is a more practical tool (than piecewise polytropes) to reach super-high densities.

 \begin{table*}
\caption{ Some neutron star properties corresponding to the red and the blue M(R) relations shown on the left-hand side of Fig.~\ref{fig_conf}. The Table is taken from Ref.~\cite{S+A_2025}.}
\label{vstable}
\centering
\begin{tabular*}{\textwidth}{@{\extracolsep{\fill}}cccccc}
\hline
\hline
$ \gamma$ & chiral order  & $M_{max}/M_{\odot}$ &   $R_{M_{max}} (km)$   & $\rho_c (fm^{-3}) $ & $R_{1.4}$ (km) \\
\hline
\hline
 3.3  & N$^2$LO & 2.19 & 10.39 &  1.09 & 11.84   \\   
        & N$^3$LO &  2.27     &  10.68   & 1.03   &  12.11   \\
3.8   & N$^2$LO & 2.43  & 11.06 &   0.93 & 12.09   \\   
        & N$^3$LO &  2.50    &  11.32   & 0.88   &  12.30   \\ 
\hline
\hline
\end{tabular*}
\end{table*}

\subsection{Subluminar and conformal}
\label{conform}

In the QCD limit of deconfined quarks in the presence of asymptotic freedom, quarks should behave like free fermions. Some perturbative QCD calculations~\cite{BWK23} support the conformal limit, $(v_s/c)^2 = 1/3$. We observed that an EoS that's subconformal at all densities (that is, Eq.~(\ref{vs}) with the asymptotic limit replaced by 1/3), cannot generate sufficiently large masses. On the other hand, the speed of sound can be asymptotically conformal and non-monotone. 
A scenario where the speed of sound peaks around few to several times nuclear density and then falls back to approach the QCD limit for deconfined quark matter would signify some sort of phase transition with the conformal limit reached well beyond central densities of the heaviest observed neutron stars. The superconformality condition, $\Big (\frac{v_s}{c} \Big )^2  > 1/3$, satisfied {\it on the average} in neutron stars may have fundamental implications on the trace of the energy-momentum tensor at the center of rotating neutron stars~\cite{Raissa+24}.

\begin{figure*}[!t] 
\centering
\hspace*{-0.1cm}
\includegraphics[width=5.6cm]{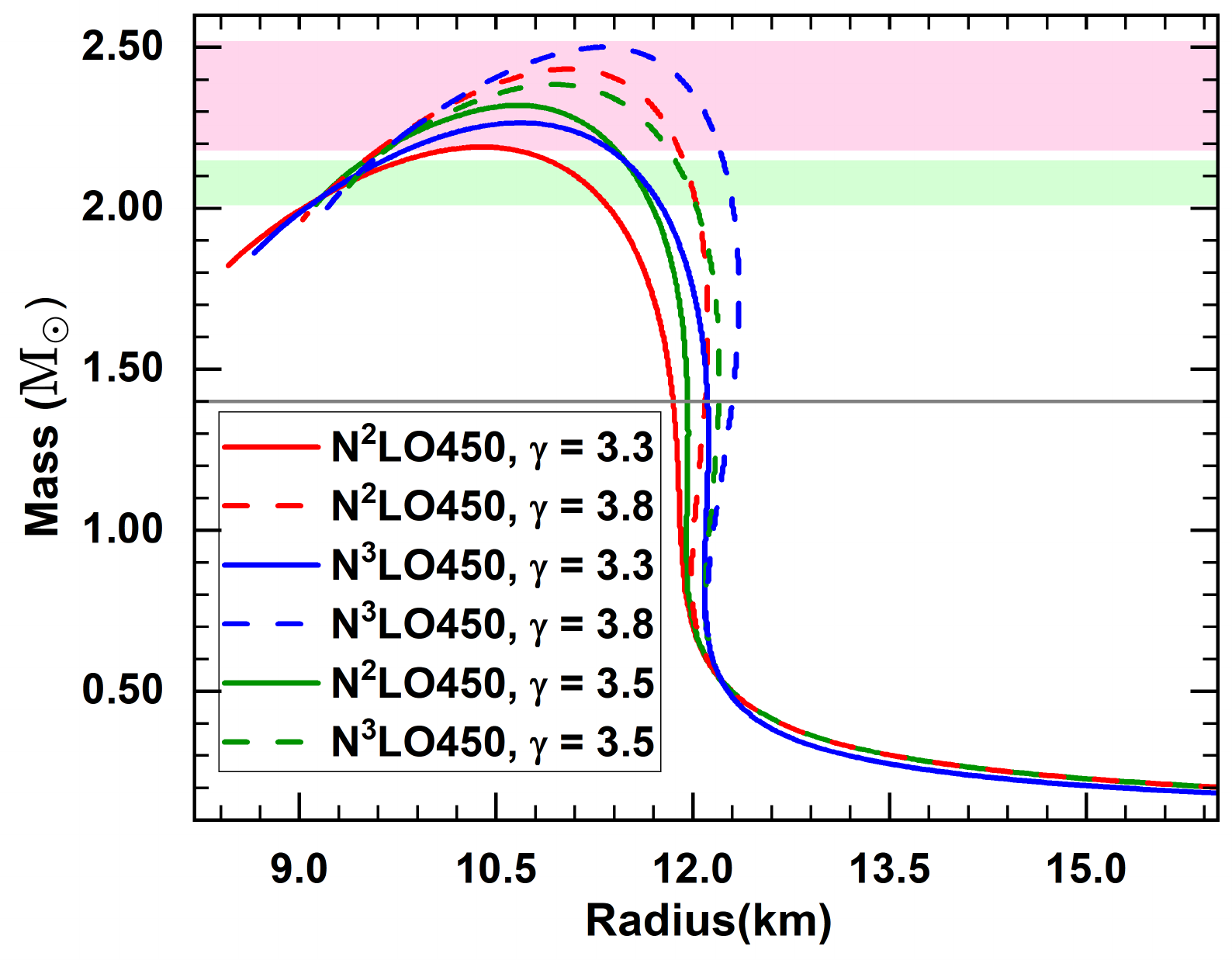}\hspace{0.01in}
\includegraphics[width=5.6cm]{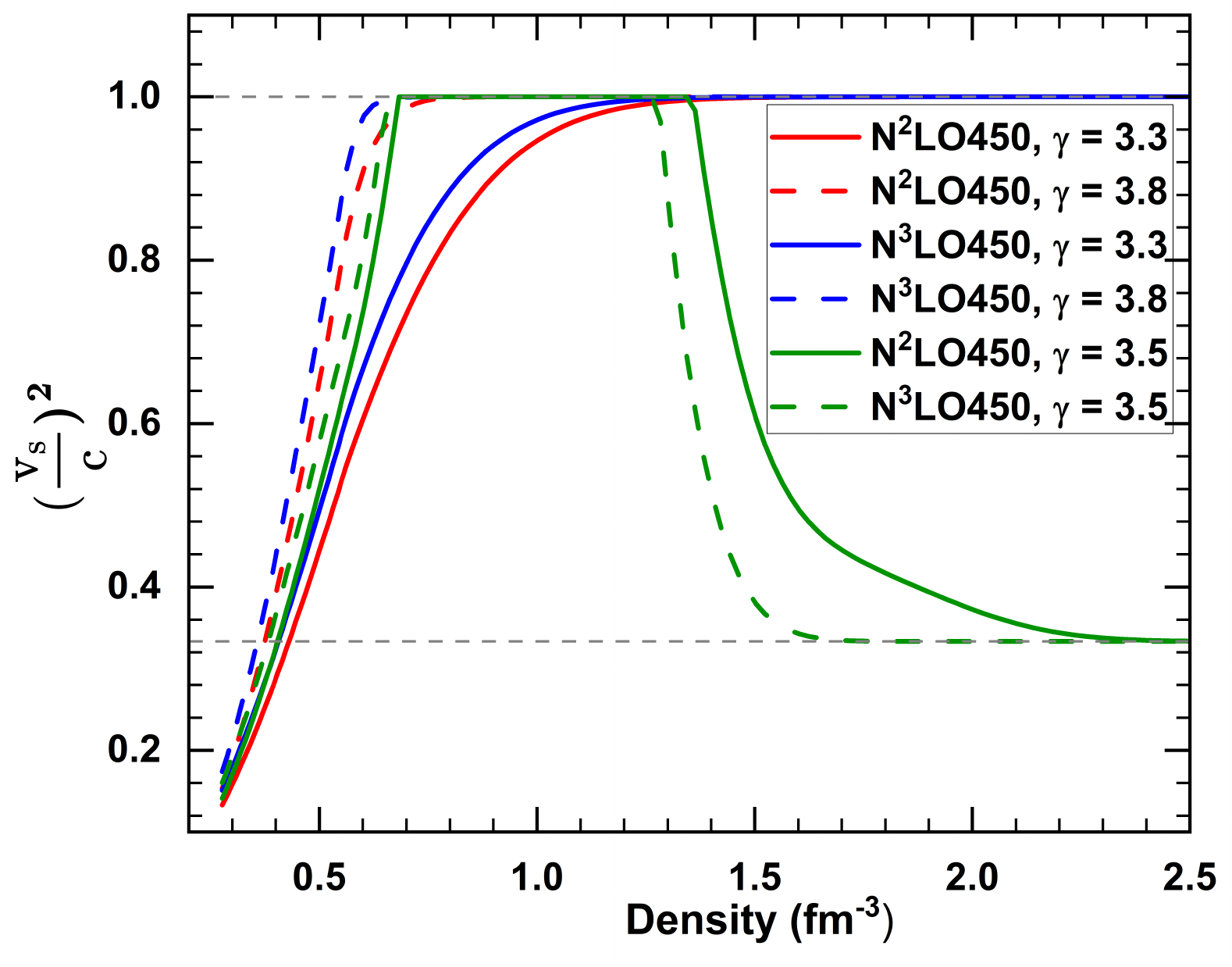}\hspace{0.01in}
\vspace*{-0.01cm}
 \caption{ Left: the red and blue M(R) curves are as described in Ref.~\cite{S+A_2025}. Green curves: M(R) relations at N$^3$LO (dashed) and at N$^2$LO (solid) where the polytropic part of the EoS has adiabatic index equal to 3.5, and the last continuation is based on the conformal speed of sound parametrization, Eq.~(\ref{skew}). 
Right: Dimensionless speed of sound squared corresponding to each of the cases displayed on the left.  See text for additional explanation.
}
\label{fig_conf}
\end{figure*}

Here, we will
 explore parametrizations where the speed of sound is consistent with the conformal limit as the density approaches infinity~\cite{Tew18}. 
As an extension of Eq.~(\ref{vs}), a natural choice might be to add a second, skewed, gaussian function, which approaches
  the asymptotic value of 1/3:
\begin{equation}
\Big (\frac{v_s}{c} \Big )^2_i = \frac{1}{3} - c_1 exp \Big [ - \frac{(\rho_i - c_2)^2}{w^2}\Big ] +
c_3 exp \Big [ - \frac{(\rho_i - c_4)^2}{w_2^2}\Big ] \Big [1 + erf \Big (c_5 \frac{(\rho_i - c_4)}{w_2} \Big ) \Big ] \; .
\label{skew}
\end{equation}
We solve for $c_1$ and $c_2$ as done above, matching Eq.~(\ref{skew}) and its derivative to the values at threshold, within a chosen parameter space for the remaining three coefficients, $c_3, c_4, c_5$. The widths $w$ and $w_2$ are obtained by normalizing both gaussians in the standard way. Although this procedure can provide a very large number of possibilities, we observed that the roots, $c_1$ and $c_2$, converge easily within limited ranges for the free parameters: $c_3 \approx [0,1]$, $c_4 \approx [1, 4] fm^{-3}$, $c_5 \approx [-50, 50]$. In fact, the roots tend to cluster around a few pairs of $c_1, c_2$ values, weakly sensitive to the large variations in $c_5$, and only moderately sensitive to variations of $c_3$ and $c_4$. As a consequence, those parametrizations produce very similar M(R) results. A representative case is shown by the green curves on the left-hand side of Fig.~\ref{fig_conf}, where we have chosen a polytropic index of 3.5 to ensure that the new curves would stand out  in between the other two sets from. The corresponding curves for the dimensionless speed of sound are seen to approach the conformal limit on the right-hand side of Fig.~\ref{fig_conf}. 
We also note that implementing the conformal limit at all densities would not allow sufficiently large masses. Maximum masses as large as those recently observed require a rapid rise of the pressure soon after the first matching point. 

It is interesting to observe that, on the left panel of Fig.~\ref{fig_conf},  the green curves are what one would expect from the non-conformal calculation and an adiabatic index that's in between the other two. In contrast, the green predictions on the right are dramatically different from the red or blue predictions. This suggests that the maximum mass is only weakly sensitive to the super-high density behavior of the EoS.

\begin{figure*}[!t] 
\centering
\hspace*{-0.1cm}
\includegraphics[width=5.6cm]{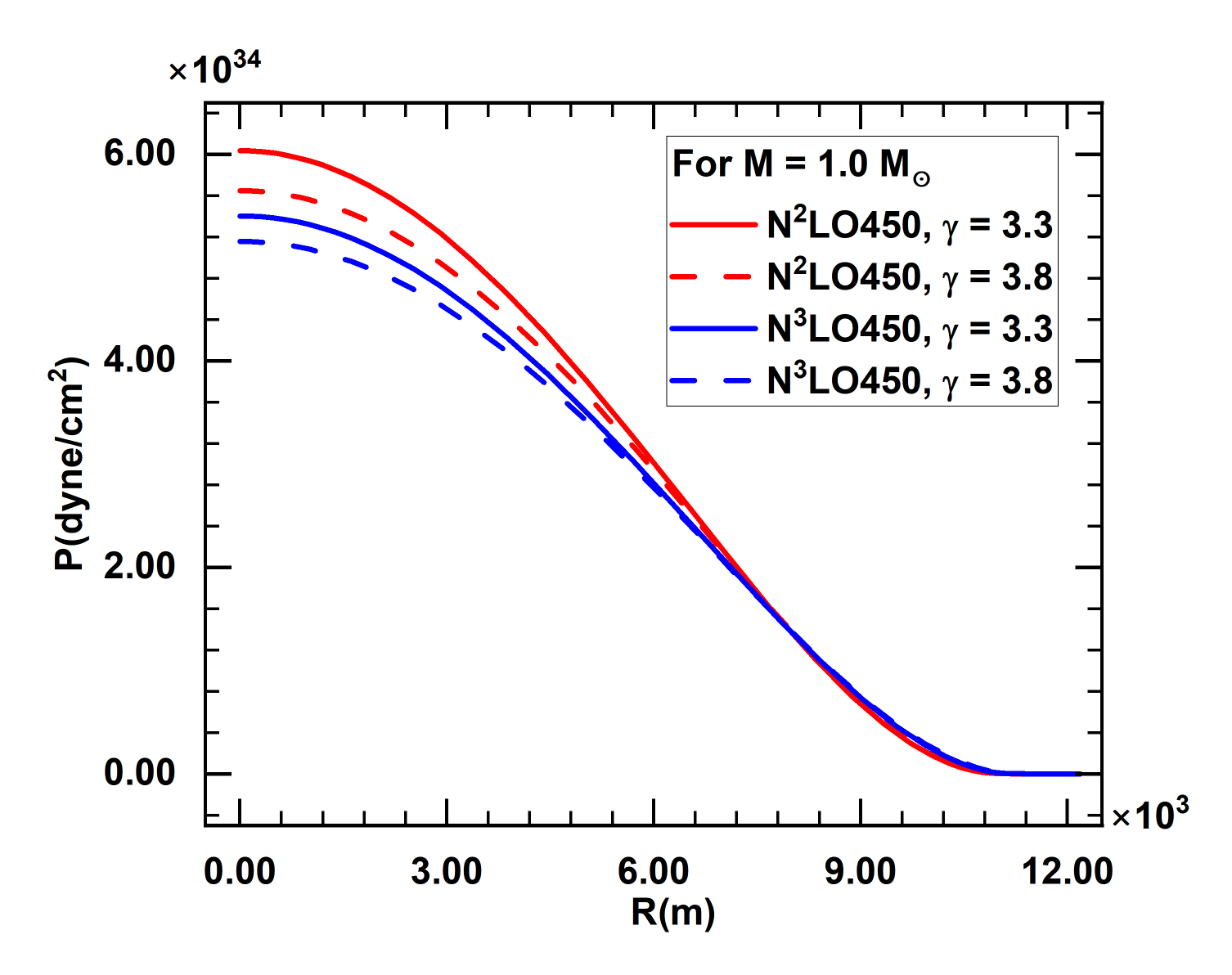}\hspace{0.01in}
\includegraphics[width=5.6cm]{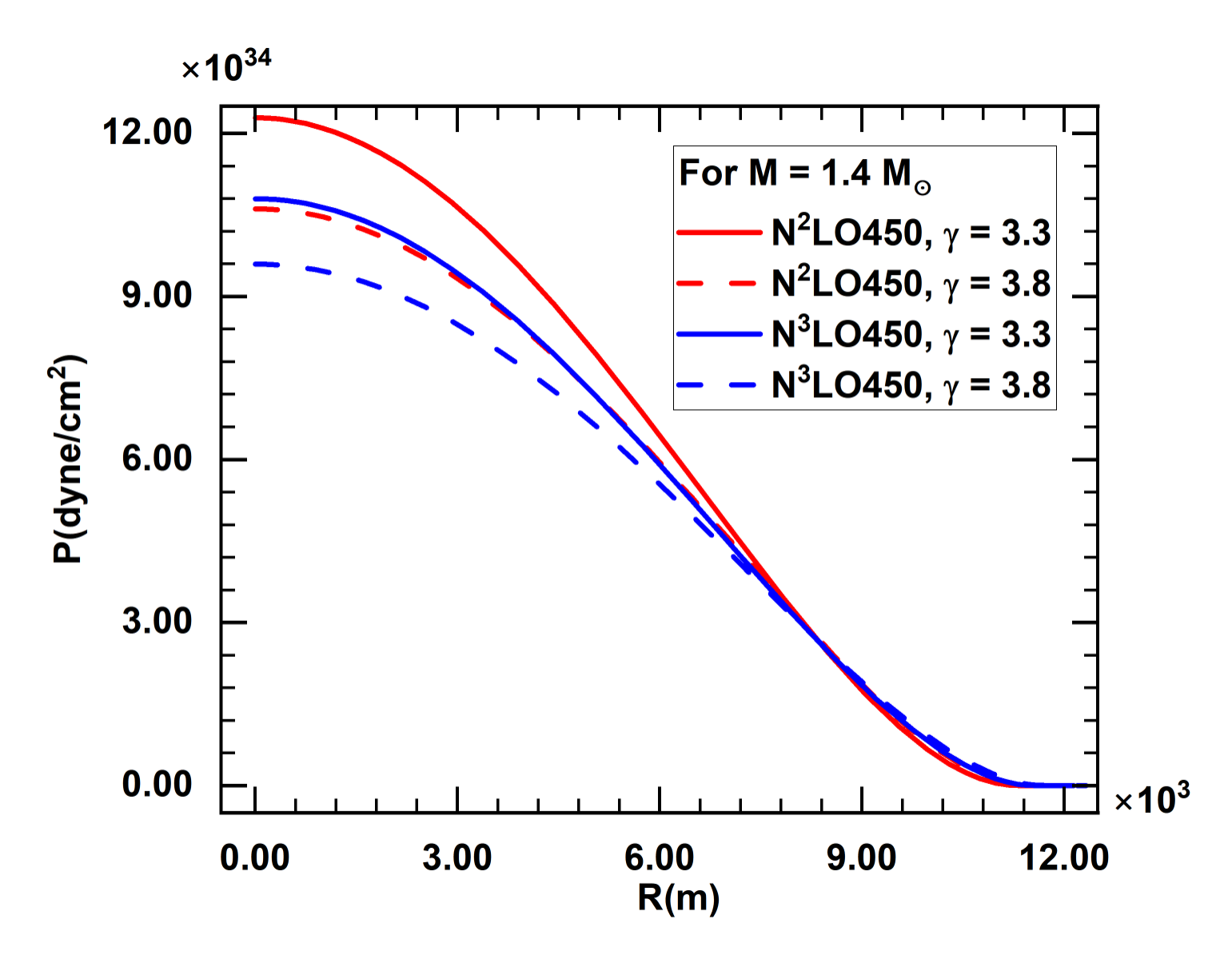}\hspace{0.01in}
\includegraphics[width=5.6cm]{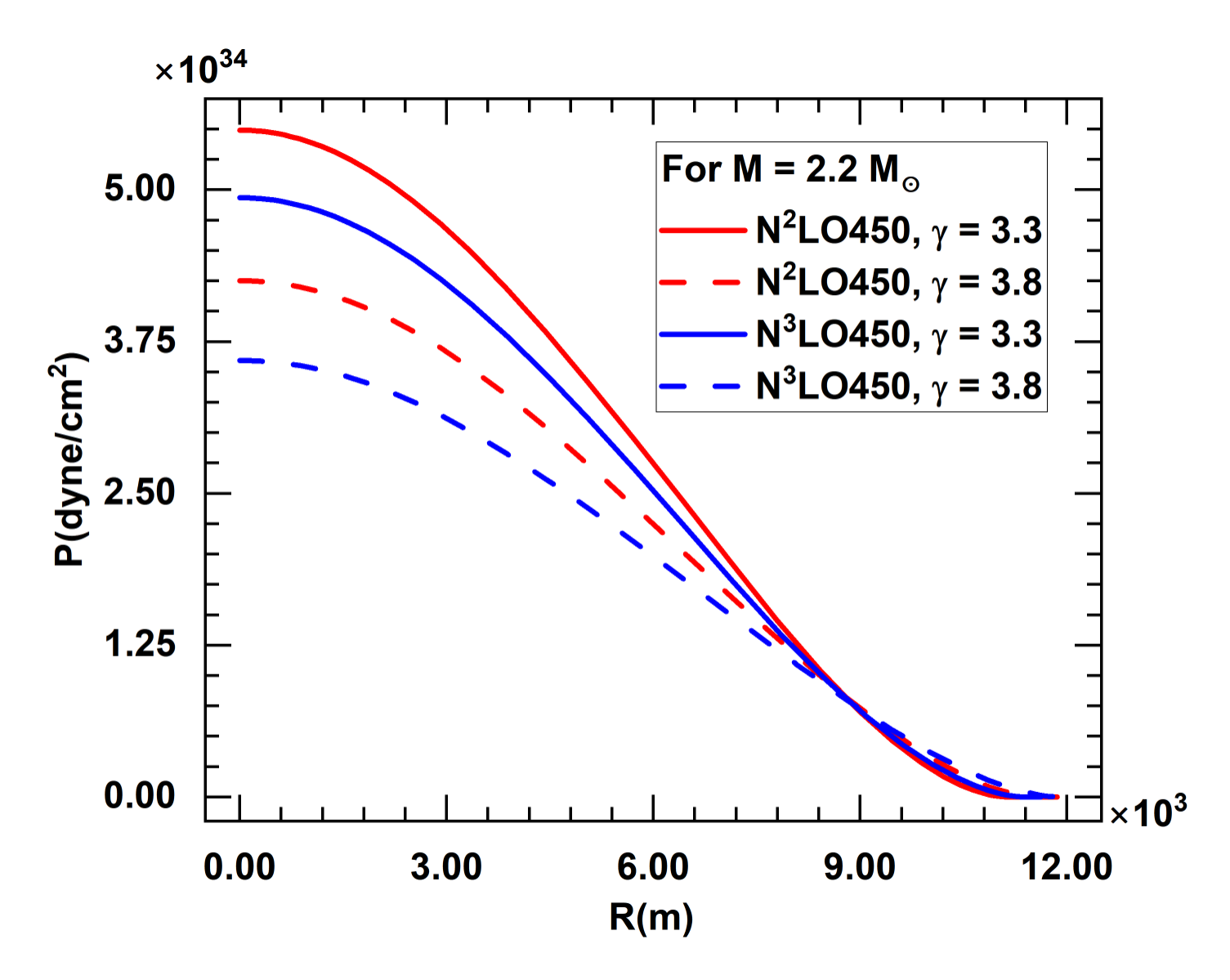}\hspace{0.01in}
\vspace*{-0.01cm}
 \caption{ Pressure profile in a low, medium, and high mass neutron star for the four cases considered in Table~\ref{vstable}.
}
\label{fig_pr}
\end{figure*}

We close this section with Fig.~\ref{fig_pr}, which displays the pressure profiles for the four EoS considered in Table~\ref{vstable}, in neutron stars with low, average, and high mass. It is insightful to compare, for each mass, the relative positions of the four curves with the corresponding ones on the left side of Fig.~\ref{fig_conf}.

\section{Tidal deformability}
\label{tide}
\subsection{General aspects}
\label{gen}

When a spherically symmetric star is placed in a quadrupolar tidal field $\epsilon_{ij}$, the response to the perturbation is a quadrupole moment, $Q_{ij}$, which, to linear order in $\epsilon_{ij}$, can be written as 
\begin{equation}
\label{QQ}
Q_{ij} = - \lambda \epsilon_{ij} \; .
\end{equation}
The tidal deformability is a parameter that quantifies how easily the star is deformed by an external tidal field. Therefore, a large tidal deformability signals a larger and less compact star which deforms easily.  Given a proposed equation of state, the tidal deformability of a neutron star of a certain mass
can be obtained by computing the metric in the asymptotic regime using the
linearized Einstein equations. 
The tidal deformability  $\lambda$ appearing in Eq.~(\ref{QQ})  is typically expressed as 
\begin{equation}
\label{lam}
 \lambda =  \frac{2}{3}k_2 R^5 \; ,
\end{equation}
where $k_2$ is the gravitational Love number, given by the expression
\begin{equation}
\nonumber
k_2(C, y_R) = \frac{8}{5} C^5 (1 - 2 C)^2 [2 - y_R + 2C(y_R-1)] \{2C[6 -3y_R + 3C(5y_R - 8)] + 
\end{equation}
\begin{equation}
\nonumber
4C^3 [13 -11 y_R + C(3y_R - 2) + 2C^2(1 + y_R)] +
\end{equation}
\begin{equation}
\label{kCR}
3(1 - 2 C)^2[2 - y_R + 2C(y_R-1)] ln(1- 2 C)\}^{-1} \; .
\end{equation}
Equation~(\ref{kCR}) is derived in the Appendix. Although analytical calculations of gravitational Love numbers and tidal deformabilities can be found in a number of excellent sources~\cite{Hind2008, Gitt+2020, Chi+2020, TC1967}, we provide here a relatively compact and self-contained derivation, for the benefit of the interested reader. We use units where c=G=1.

Of greater astrophysical relevance is the dimensionless tidal deformability, defined as
\begin{equation}
\label{Lam}
 \Lambda =  \frac{\lambda}{M^5} = \frac{2}{3}k_2 C^{-5} \; ,
\end{equation}
 where C is the neutron star's compactness, C=M/R.  It is $\Lambda$ that actually governs the effects of tides on inspiral waveforms~\cite{VanOe+2017}.

\begin{figure*}[!t] 
\centering
\hspace*{-0.1cm}
\includegraphics[width=5.5cm]{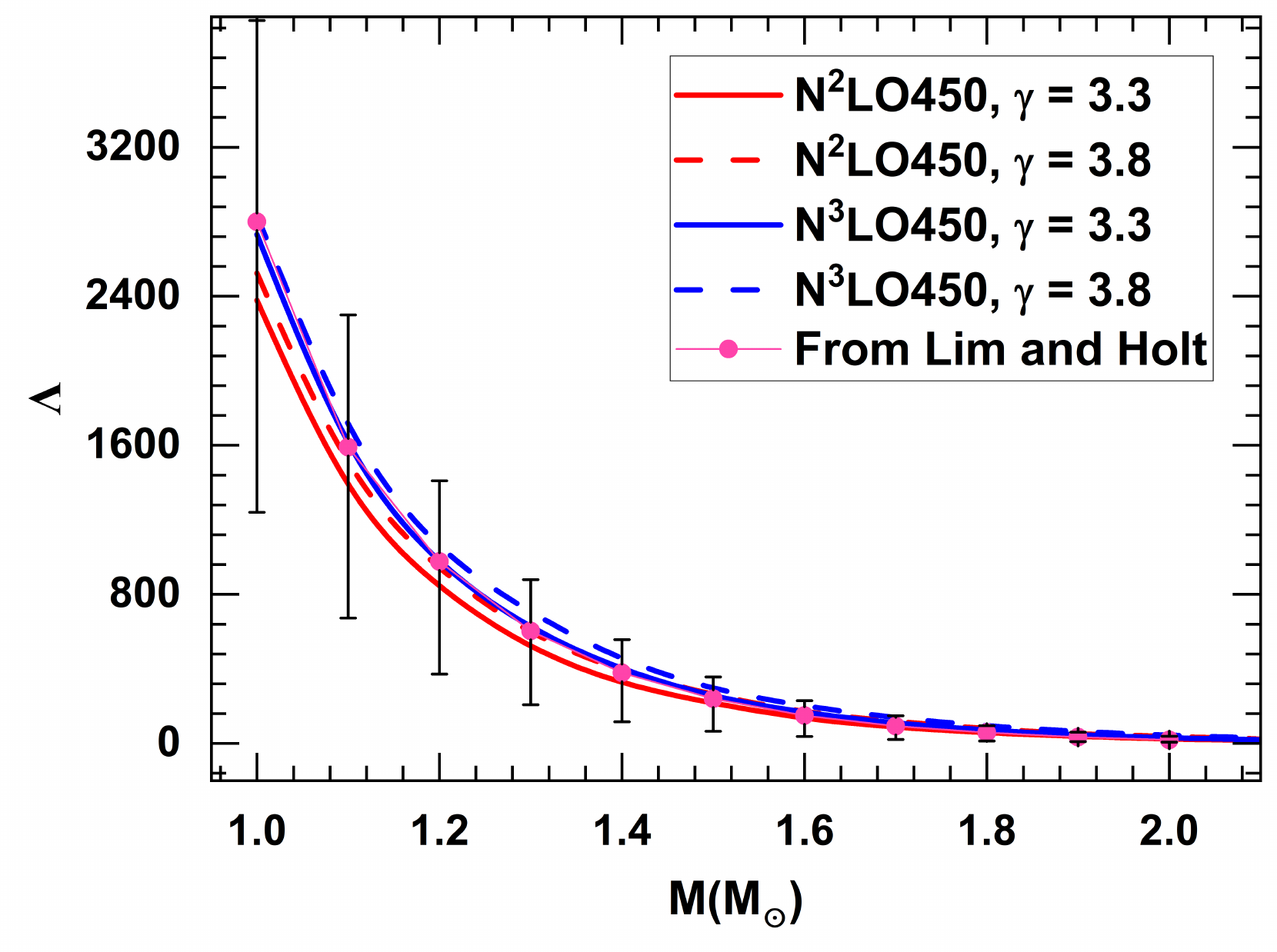}\hspace{0.01in}
\includegraphics[width=5.9cm]{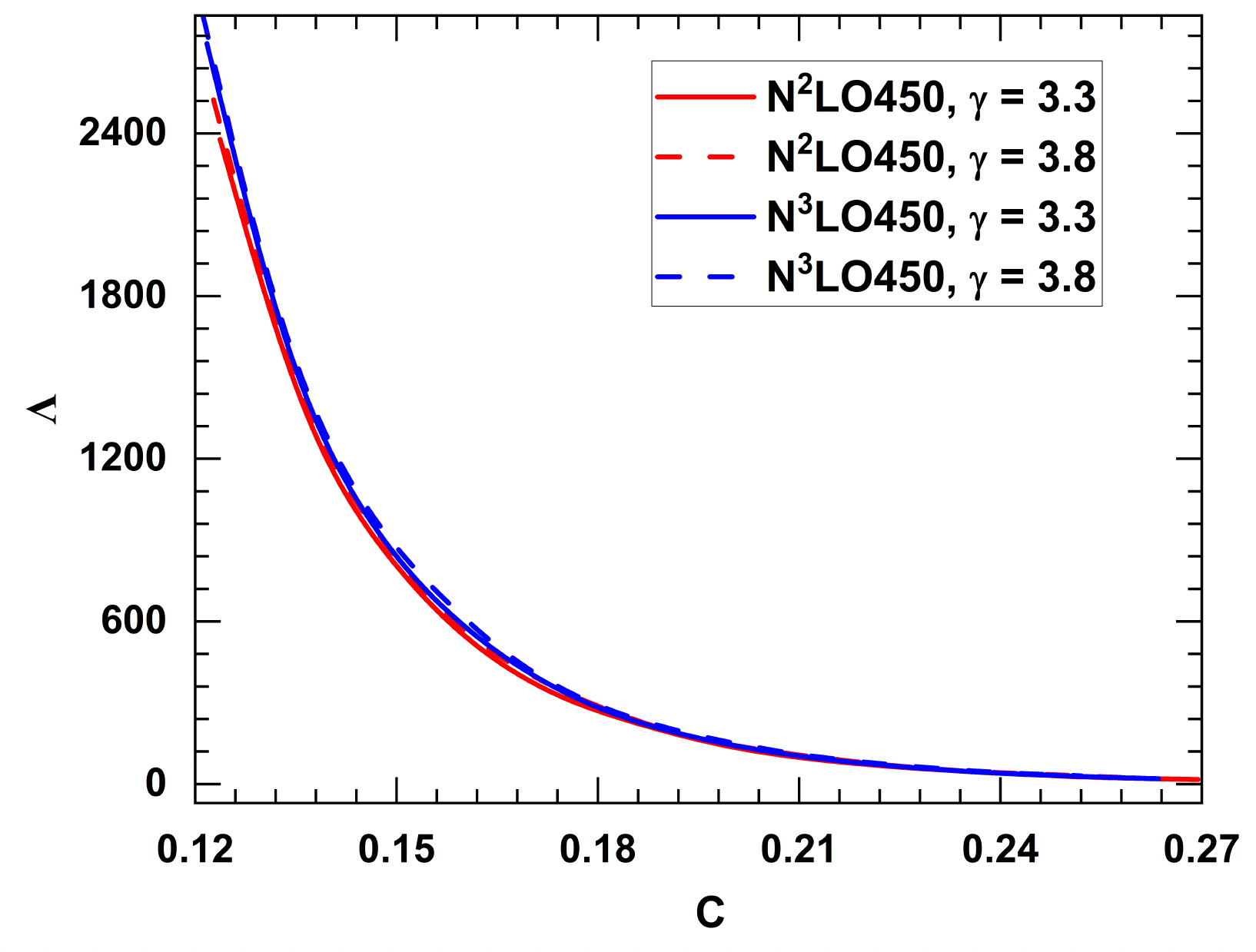}\hspace{0.01in}
\vspace*{-0.01cm}
 \caption{ Left: The dimensionless tidal deformability as a function of the neutron star mass (left) and as a function of the compactness (right). The pink dots and their uncertainties are taken from the Bayesian analysis of Ref.~\cite{LH19}. See text for more details. }
\label{lambda_vs_m_C}
\end{figure*}

\begin{figure*}[!t] 
\centering
\hspace*{-0.2cm}
\includegraphics[width=5.5cm]{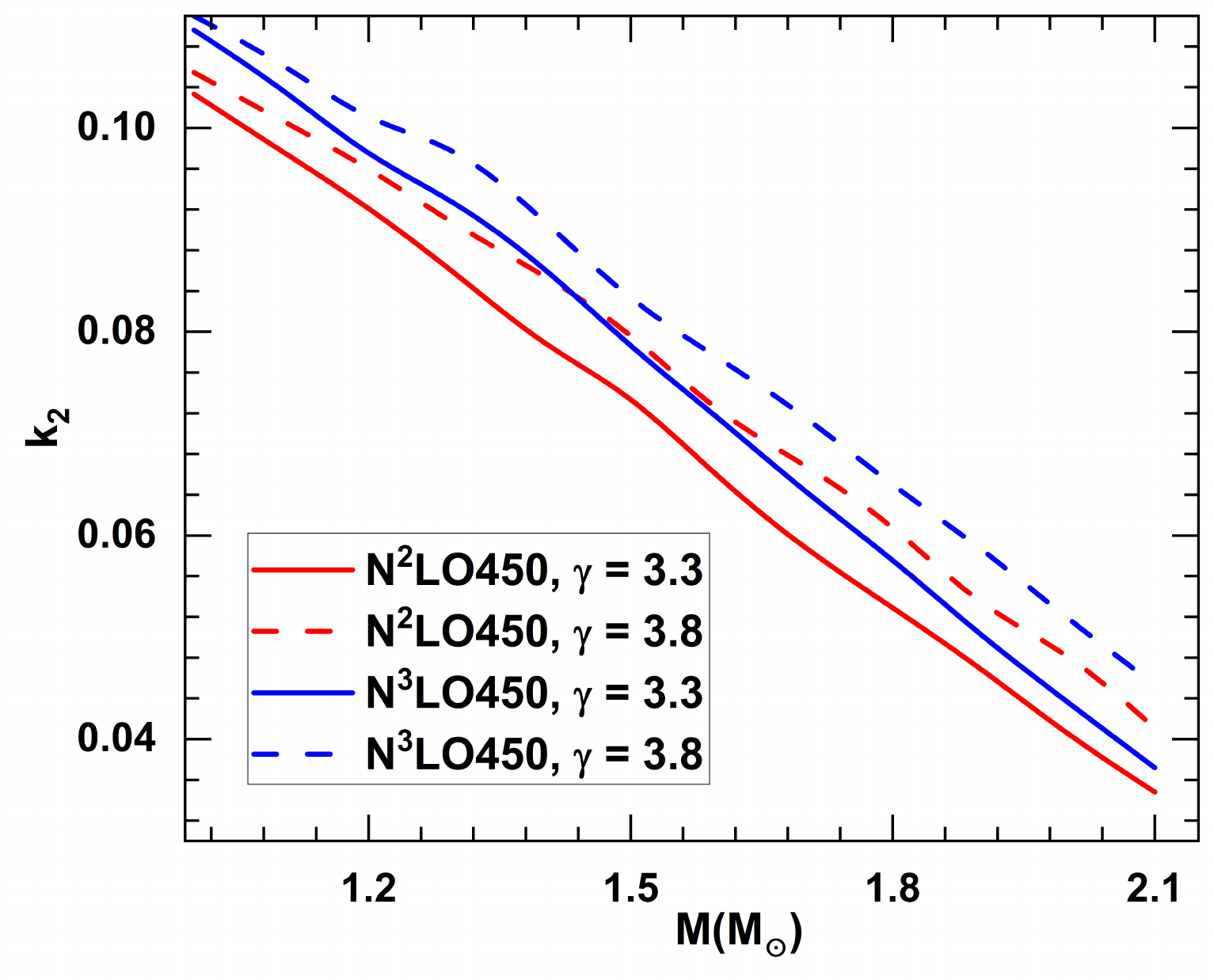}\hspace{0.01in}
\includegraphics[width=5.5cm]{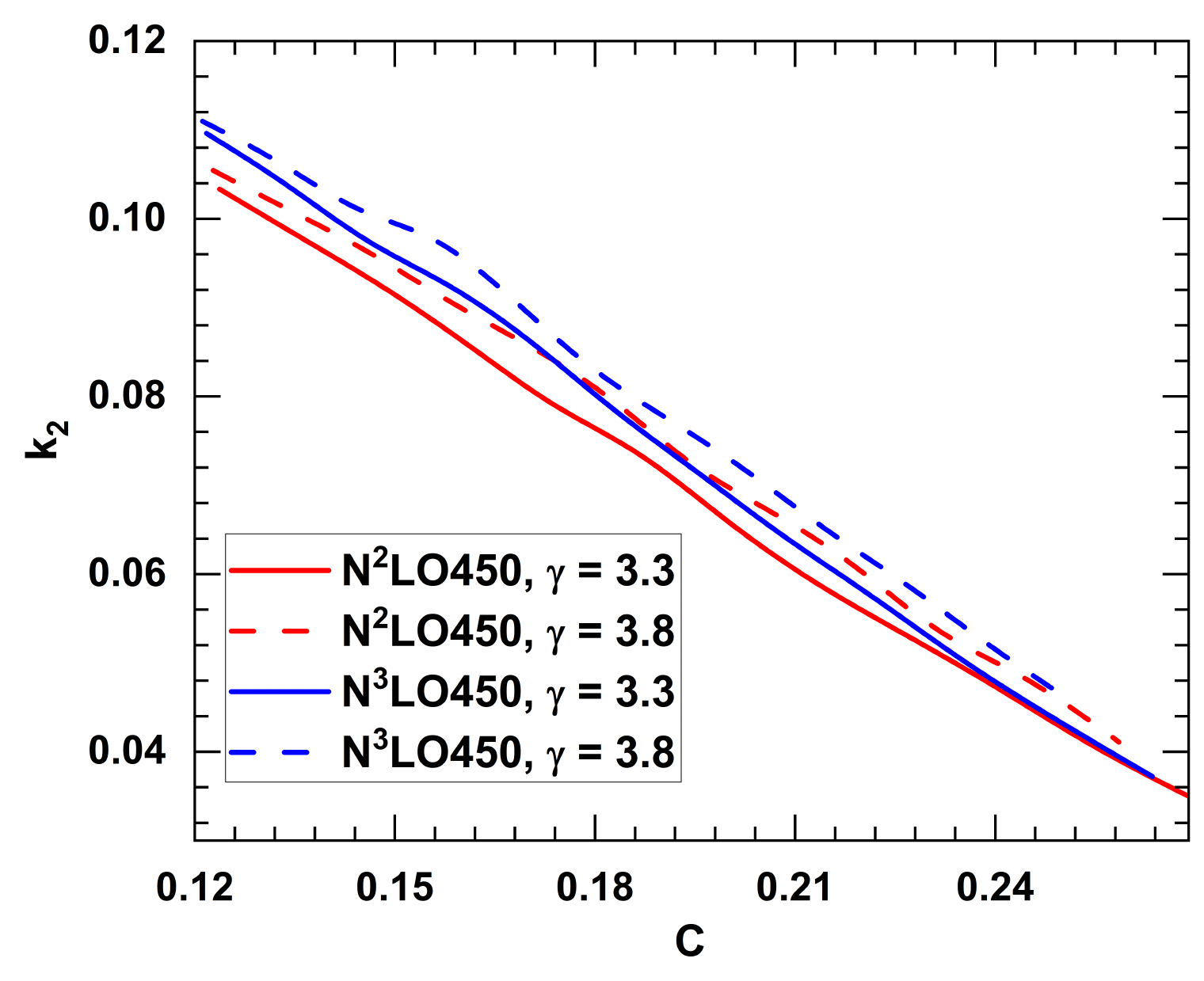}\hspace{0.01in}
\vspace*{-0.01cm}
 \caption{ Left: The tidal Love number as a function of the neutron star mass (left) and as a function of the compactness (right). }
\label{k2_vs_m_C}
\end{figure*}

\subsection{Results and discussion}
\label{res}
\subsubsection{Tidal deformability and tidal Love number}

On the left side of Fig.~\ref{lambda_vs_m_C}, we show the dimensionless tidal deformability for the four cases considered in Table~\ref{vstable}. For comparison, the pink dots are taken from the Bayesian analysis of Ref.~\cite{LH19} and represent the most probable values from sampling 300,000 EoS models based on chiral EFT and experimental constraints. The vertical bars show the uncertainties at the 2$\sigma$ level. All of our predictions are close to each other and quite consistent with the most probable outcomes from the Bayesian analysis. On the right of the figure, our predictions are displayed {\it versus} the star's compactness.  

Closely related to $\Lambda$ is the tidal Love number $k_2$, related to $\Lambda$ as in Eq.~(\ref{Lam}). This is shown in Fig.~\ref{k2_vs_m_C} as a function of the neutron star mass and compactness. Often, comparison between relativistic and Newtonian calculations or tabulations of deformation properties for polytropic EoS with different polytropic indices are given in terms of $k_2$ rather than $\Lambda$~\cite{Hind2008}. Our predicted values of $k_2$ are shown in Table~\ref{k2} for a representative case. 

\begin{table*}
\caption{Compactness and tidal Love number as a function of the neutron star mass for one of the EoS used in Fig.~\ref{k2_vs_m_C}.}
\label{k2}
\centering
\begin{tabular*}{\textwidth}{@{\extracolsep{\fill}}cccc}
\hline
\hline
 EoS &                                     $M/M_{\odot}$ &   C   & $ k_2$  \\
\hline
\hline
 N$^2$LO  $\gamma$=3.3        & 0.90        &     0.112      & 0.105 \\
                                                 & 1.0        &      0.125   &   0.103   \\   
                                                 & 1.1        &       0.137  &   0.0972    \\
                                                &  1.2        &     0.150   &   0.0926    \\   
                                                & 1.3        &      0.163   &   0.0853   \\   
                                                  & 1.4      &       0.175  &    0.0782 \\
                                                 & 1.5       &      0.188  &      0.0722 \\
                                                &  1.6       &       0.202 &      0.0660 \\
                                                &   1.7      &       0.215  &      0.0579   \\
                                                 &   1.8     &        0.230   &    0.0521 \\
                                               & 1.9        &         0.245    &   0.0428 \\
                                               &  2.0       &        0.261    &    0.0361 \\
                                                &  2.2      &       0.313   &   0.0161    \\   
\hline
\hline
\end{tabular*}
\end{table*}

Recent multimessanger constraints can be found in Ref.~\cite{Huang25}, obtained from various methodologies based on variants of the GW170817 data sets. Their most robust constraint is based on a combination of GW-based findings and EoS-independent X-ray constraints, and is reported as
\begin{equation}
\label{constr}
\Lambda_{1.4} = 265.18^{+237.88}_{-104.38} \; \; \; \; \; \; \; \;   R_{1.4} = 11.53^{+0.89}_{-0.88} \; \mbox{km} \; ,
\end{equation}
for the dimensionless tidal deformability and radius of the canonical mass neutron star.  We underline that this EoS-independent constraint is yet another confirmation that the large radius extracted from the PREX neutron skin measurement~\cite{Ree21} is unrealistic.

We account for both chiral truncation error (cf. Sec.~\ref{TF}), and the spreading due to different values of $\gamma$. Combining the two independent uncertainties in quadrature, we estimate our ranges for $\Lambda_{1.4}$ and $R_{1.4}$ at N$^2$LO to be
\begin{equation}
\Lambda_{1.4} = 355.25 \pm 85.64 \; \; \; \; \; \; \; \;   R_{1.4} = (11.99 \pm 0.51) \; \mbox{km} \; ,
\label{LR}
\end{equation}
quite consistent with Eq.~(\ref{constr}). 
We underline that the overall uncertainties in Eq.~(\ref{LR}) are  evaluated at N$^2$LO, which, at this time, is a fully consistent order, see comments in Sec.~\ref{TF}.

Other inferred
upper limits on neutron star radii and tidal deformabilities
from GW170817 rule out stiff equations of state that result
in large radii and large tidal deformabilities. Recent estimates of the measurability of tidal effects and the ability of
these observatories to constrain the EoS with signals from black hole-neutron star and  binary neutron star systems can be found in
 Refs.~\cite{Hot+2016, Dam+2012, Del+2013, Aga+2015, Wade+2014} and references therein.

Next, we want to explore the sensitivity of $\Lambda$ to the high-density parametrization of the speed of sound discussed in Sec.~\ref{conform}. For that purpose, we employ the four EoS used for Fig.~\ref{fig_conf}, keeping the same color and pattern conventions. As for the M(R) results on the left side of Fig.~\ref{fig_conf}, we see that the green curves are what we would expect from the non-conformal calculation and an adiabatic index that's between the other two, while the speed of sound predictions with or without the conformal limit are dramatically different, see right side of Fig.~\ref{fig_conf}. This seems to indicate that the tidal deformability is mostly sensitive to the EoS at medium to moderately high densities, a regime that ``feels" the influence of the microscopic theory~\cite{S+A_2025}, as discussed in Sec.~\ref{non_conf}. 

\begin{figure*}[!t] 
\centering
\hspace*{-0.1cm}
\includegraphics[width=5.5cm]{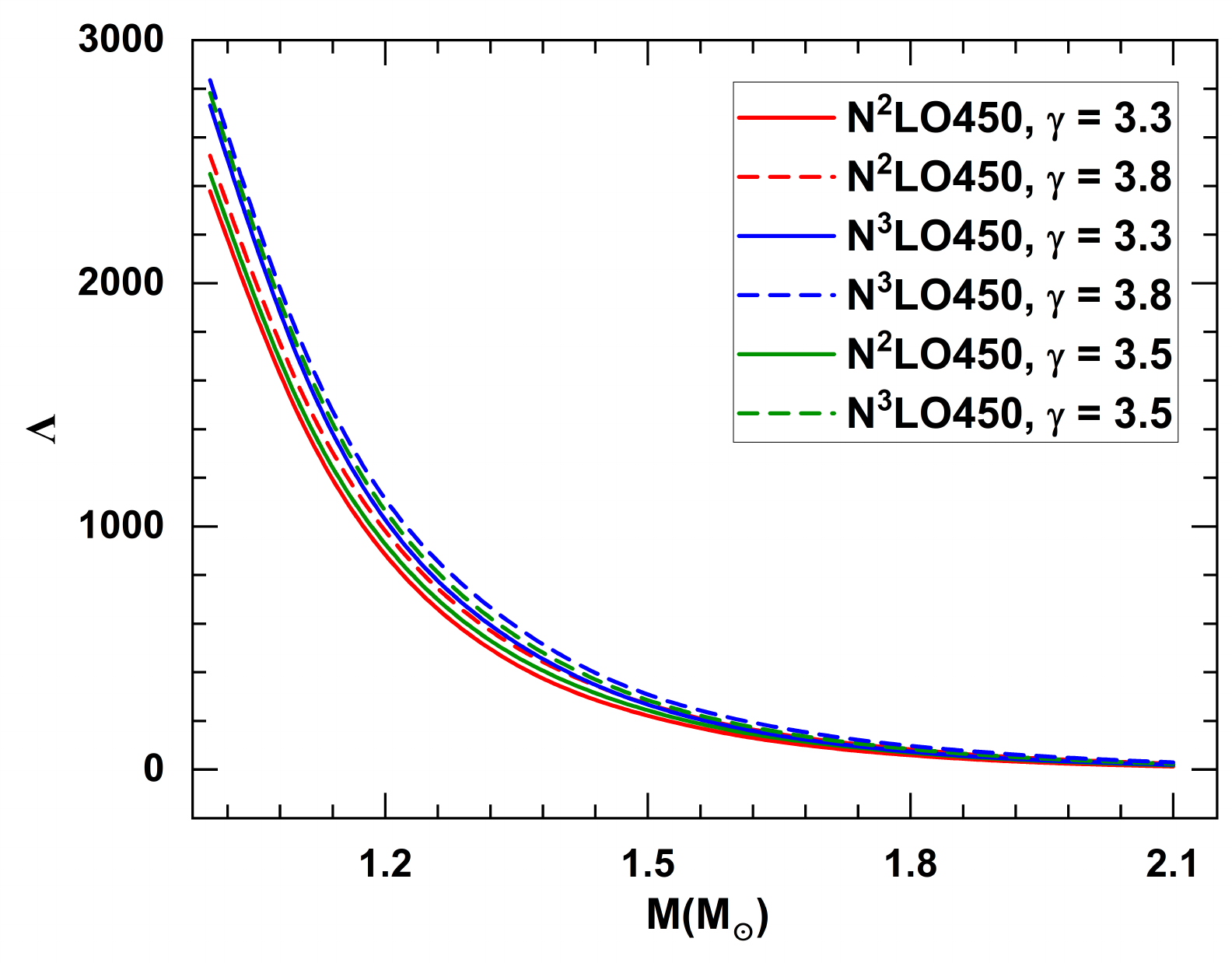}\hspace{0.01in}
\includegraphics[width=5.5cm]{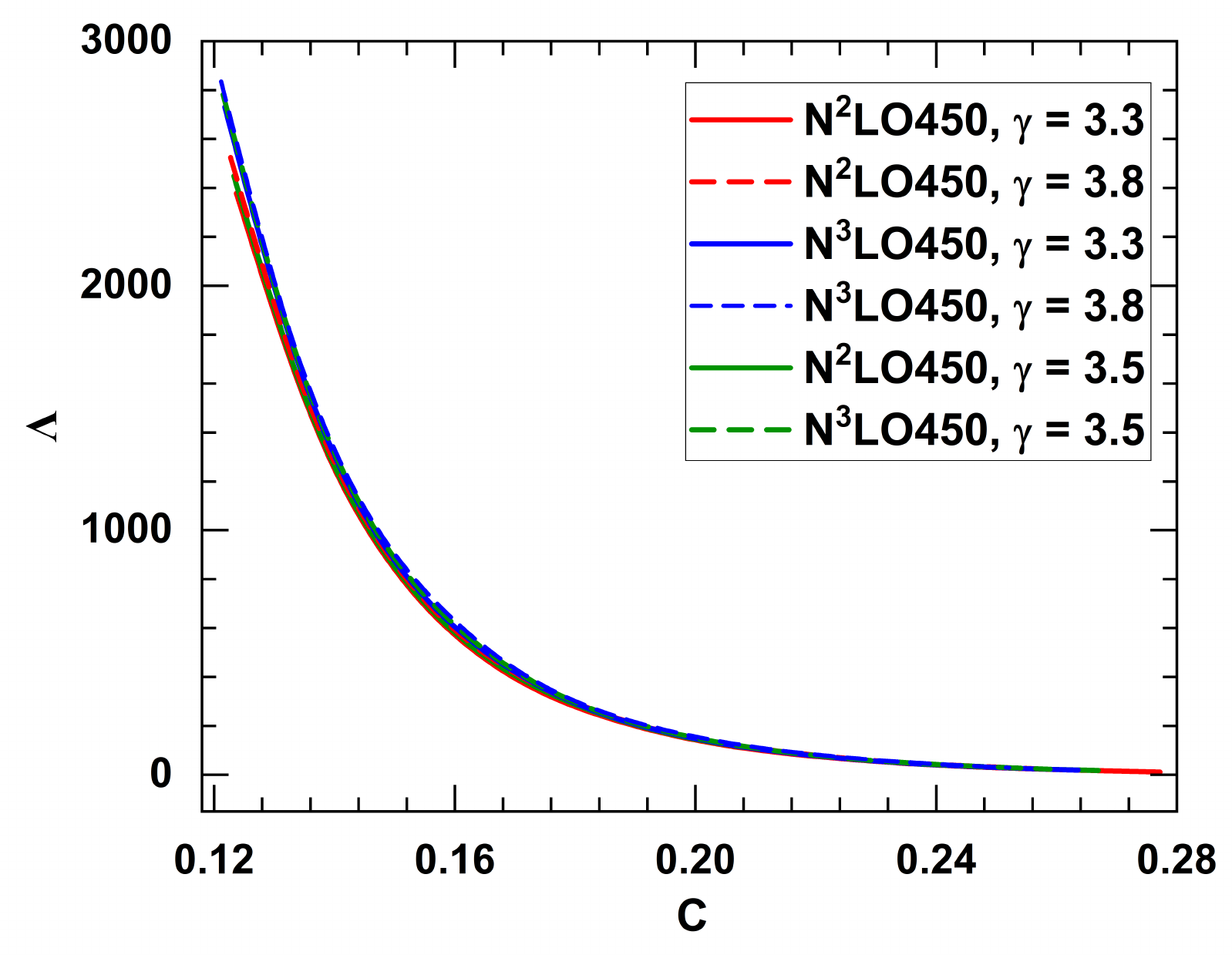}\hspace{0.01in}
\vspace*{-0.01cm}
 \caption{Tidal deformability as a function of mass (left) and compactness (right). The green curves are based on the EoS extended to the highest densities with a conformal parametrization of the speed of sound. }
\label{green}
\end{figure*}   
 
To make the above statement more quantitative, next we perform tests adopting different extension methods. Prior to exploring other options, our typical high-density extension consisted of piecewise polytropes with appropriately chosen adiabatic index. Equations of state unable to support a maximum mass of at least 2.01 M$_{\odot}$ were discarded, and M(R) relations were cut at the central density where causality was violated. With these guidelines, we selected a few favorable combinations, given in Table~\ref{gama}.

\begin{table*}
\caption{Maximum mass and $R_{1.4}$ generated by EoS continuations in terms of two polytropes with index $\Gamma_1$ and $\Gamma_2$, respectively. Values taken from Ref.~\cite{S+A_2025}.}
\label{gama}
\centering
\begin{tabular*}{\textwidth}{@{\extracolsep{\fill}}ccccc}
\hline
\hline
 EoS label & $\Gamma_1$ &   $\Gamma_2$ & $M_{max}/M_{\odot}$ & $R_{1.4}$ (km) \\
\hline
\hline
 A &3.1 & 2.7&  2.10 & 12.00   \\   
 B &3.1  & 2.8 & 2.12 &12.00 \\
 C & 3.2  & 2.7 & 2.15 & 12.06 \\
 D & 3.2  &  2.8 &2.17 & 12.06 \\
 E & 3.3 & 2.7 & 2.19 & 12.11 \\
\hline
\hline
\end{tabular*}
\end{table*} 
In Table~\ref{high-rho}, we will compare Love numbers and $\Lambda_{1.4}$ for diverse high-density continuations.

\begin{table*}
\caption{Tidal Love number and deformability of the canonical-mass neutron star, obtained with the described EoS. The labels A through E correspond to the same labels in Table~\ref{gama}.  }
\label{high-rho}
\centering
\begin{tabular*}{\textwidth}{@{\extracolsep{\fill}}ccc}
\hline
\hline
 EoS description & $k_2$   & $\Lambda_{1.4}$ \\
\hline
\hline
 A &  0.083812 &  373.908325 \\   
 B &  0.083812  & 373.908325  \\
 C & 0.085187 & 388.95812 \\
 D &  0.085187 & 388.95812 \\
 E &  0.086453 & 402.891215 \\
\hline
N$^3$LO, $\gamma$=3.3 &   0.086512    &   403.520110    \\
 $(v_s/c)^2$ $\rightarrow$ 1   &        &      \\
 N$^3$LO, $\gamma$=3.8 &   0.091063    &      457.909912   \\
  $(v_s/c)^2$ $\rightarrow$ 1&         &          \\
\hline
 N$^3$LO, $\gamma$=3.3 &   0.086512    &   403.523460    \\
 ($v_s/c)^2$ $\rightarrow$1/3   &       &  \\
N$^3$LO450, $\gamma$ = 3.8         &     0.091064       &   457.911549   \\
 ($v_s/c)^2$ $\rightarrow$ 1/3   &              &               \\
\hline
\hline
\end{tabular*}
\end{table*} 
For the first group in Table~\ref{high-rho}, above the first horizontal line, the mild, if any, sensitivity to the high-density continuation is perhaps not surprising, since EoS A through E are not fundamentally different. The second group highlights the sensitivity to the polytropic index employed at medium-high densities. Finally, the third and last group, together with the second, highlights the sensitivity to the super-high density parametrization (for equal $\gamma$).
We conclude that, in all of the above cases, differences are essentially negligible compared to the current uncertainties for these quantities. In particular, the last two entries in Table~\ref{high-rho} demonstrate that the super-high density continuation has no impact on these observables. 

\subsubsection{The mass-weighted tidal deformability}

Here, we focus on our results for the effective tidal deformability, $\tilde{\Lambda}$, for a binary with masses (deformabilities) $M_1({\Lambda_1})$ and $M_2({\Lambda_2})$.  First, we recall that $\Lambda_1$ or $\Lambda_2$ cannot be disentangled in the observed gravitational wave, and thus are not directly observable. Instead, it is possible to constrain the mass-weighted average (or effective) tidal deformability,
\begin{equation}
\tilde{\Lambda} = \frac{16}{13} \frac{(M_1 + 12 M_2)M_1^4 \Lambda_1 +  (M_2 + 12 M_1)M_2^4 \Lambda_2} {(M_1 + M_2)^5} \; ,
\label{eff}
\end{equation}
with $M_1$ and $M_2$ the masses of the primary and the secondary neutron stars in the binary. $\tilde{\Lambda}$ determines the total effect of the tidal deformability on the phase evolution of the GW signal.
Another observable that can be extracted from the evolution of the frequency in the observed GW signal is
 the chirp mass, defined as 
\begin{equation}
M_c = \frac{(M_1 M_2)^{3/5}}{(M_1 + M_2)^{1/5}} = M_1 \frac{q^{3/5}}{(1 + q)^{1/5}} \; ,
\label{chirpM}
\end{equation}
with $q= M_1/M_2$.
 The chirp mass is one of the most tightly constrained properties extracted from GW170817. Therefore, $M_c$ can be set equal to the observed value for the purpose of comparing predictions of $\tilde{\Lambda}$ with experimental data.
The GW170817 constraint was stated as $ M_c/M_{\odot} = 1.188^{+0.004}_{-0.002} $ at the 90\% confidence level~\cite{Abb17a}, from which the primary and secondary masses (in units of $M_{\odot}$), were inferred to be within the range (1.36, 1.60) and (1.17, 1.36), respectively, placing $q$ between 0.73 and 1. We exploited this information to calculate the values in Table~\ref{chirp}. Starting with $M_1$ in the full range of masses inferred from GW170817, (1.17, 1.60), and imposing the very stringent $M_c$ constraint, we obtain $\tilde{\Lambda}$ for each value of $q$, see Table~\ref{chirp}. The effective tidal deformability changes by about 4\% across the $q$-variations. Essentially, $\tilde{\Lambda}$ is constrained at fixed mass.

\begin{table*}
\caption{Effective tidal deformability under the GW170817 conditions.}
\label{chirp}
\centering
\begin{tabular*}{\textwidth}{@{\extracolsep{\fill}}cccc}
\hline
\hline
 $q=M_2/M_1$         &  $M_1/M_{\odot} $        &   $M_2/M_{\odot} $        & $\tilde{\Lambda}$ \\
\hline
\hline
    0.73    &    (1.599,1.600)             & (1.167,1.168)                &  (390.32, 388.61)  \\   
       0.75    &    (1.577,1.584)             & (1.183,1.188)                &  (393.76, 384.96)  \\         
 0.80    &    (1.526, 1.532)             & (1.221,1.226)                &  (399.19, 389.61[)  \\              
0.85    &    (1.479, 1.486)             & (1.257, 1.263)                &  (401.30, 380.37)  \\        
0.90    &  (1.437, 1.443)      &         (1.293, 1.299)                 & (392.83, 382.89) \\
0.95    &   (1.398, 1.404) &            (1.328, 1.334)                   &(396.11, 383.23) \\
 1.0     &   (1.363, 1.369)  &            (1.363, 1.369)   &              (393.71, 383.98)    \\                    
\hline
\hline
\end{tabular*}
\end{table*}

\begin{figure*}[!t] 
\centering
\hspace*{-0.1cm}
\includegraphics[width=7.0cm]{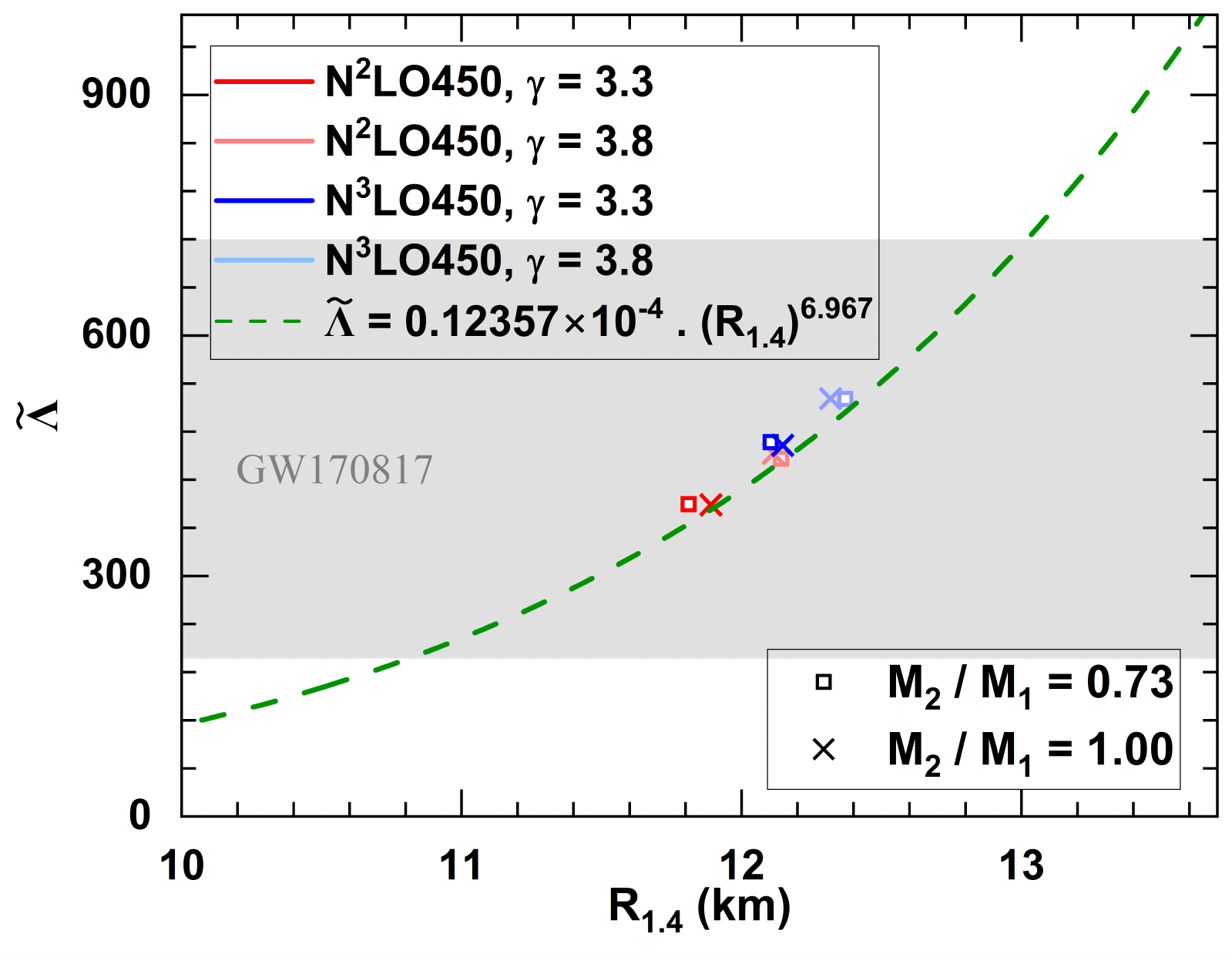}\hspace{0.01in}
\vspace*{-0.01cm}
 \caption{Effective deformability vs. the radius of the canonical mass neutron star. The shaded area represents constraints from GW170817 and AT2017gfo~\cite{Cough+18}. The continuous curve is a parametrization of $\tilde{\Lambda}$ vs. $R_{1.4}$~\cite{Kol+21}. Our predictions are indicated by the squares (q=0.73) and the crosses (q=1.0). }
\label{tilde_R}
\end{figure*}

As we already pointed out when discussing $\Lambda_{1.4}$, 
one of the most physically interesting aspect of the GW170817 data is the opportunity to extract constraints on the radius of the canonical mass neutron star and thus discriminate among various EoS. In Fig.~\ref{tilde_R}, the shaded area represents constraints from GW170817 and AT2017gfo~\cite{Cough+18}, which confine $\tilde{\Lambda}$ to the interval (197, 720). The curve is a typical power law parametrization of $\tilde{\Lambda}$  as a function of $R_{1.4}$, see Ref.~\cite{Kol+21} and references therein. Clearly, radii smaller than 10.8 km and larger than about 13 km are ruled out.

\subsubsection{Additional discussion}
\label{terr}

Our predictions are generally consistent with multimessenger constraints, as well as terrestrial laboratory constraints. For the purpose of rendering the paper more selfcontained, this short section will address recent terrestrial constraints.

 For the symmetry energy, its density dependence, and the pressure in NM, {\it vis a vis} constraints from Ref.~\cite{LT22}, the reader is referred to Ref.~\cite{Sam23}. Reference~\cite{Sam24} presents a broad analysis of microscopic predictions {\it versus} empirical constraints for the neutron skin of $^{48}$Ca and $^{208}$Pb. In that work, we emphasize that the small skin of $^{48}$Ca extracted from CREX is quite consistent with microscopic predictions and available experimental constraints, see Fig.1 of Ref.~\cite{Sam23} and Ref.~\cite{Sam22}.

 The statistical analysis of Ref.~\cite{Koehn+25} employed 100,000 candidate EoS to obtain Bayesian posteriors on the symmetry energy parameters at normal density. The authors reanalyzed the results of PREX and CREX and concluded that neutron star radius posteriors can be statistically compatible with PREX and CREX. They discuss
 the question of how each constraint on the EoS
is constructed from the observed data, and the various assumptions that
could introduce biases. They comment
on potential sources of systematic errors while applying a
particular constraint, and discuss how different statistical interpretations of the data may affect the
inferred results.
The authors conclude ``... {\it the need 
for enhanced theoretical models in order to reduce the
systematic uncertainties currently present and possibly
accommodate both the PREX-II and CREX results}." 
We take the opportunity to add that theory improvements should never be directed at accomodating specific constraints.

\begin{figure*}[!t] 
\centering
\hspace*{-1cm}
\includegraphics[width=6.0cm]{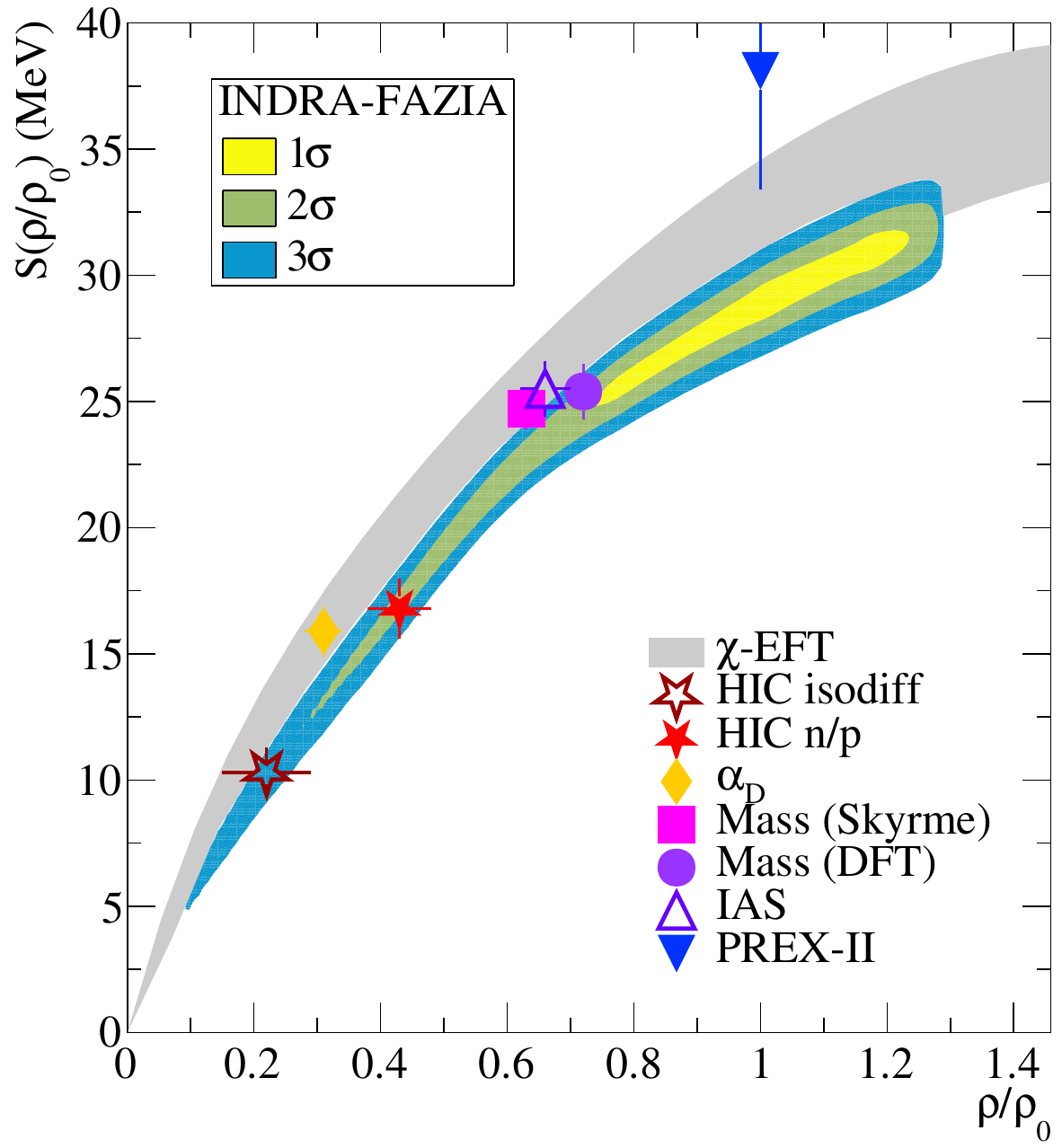}\hspace{0.01in} 
\vspace*{-0.1cm}
 \caption{(Color online) The figure was obtained from one of the authors Ref.~\cite{Ciampi+25} under the CC BY license  http://creativecommons.org/licenses/by/4.0/. The yellow, green, and blue areas are the constraints at the 1$\sigma$, 2$\sigma$, and 3$\sigma$ confidence levels, respectively. The markers signify constraints
extracted from other HIC experiments or different methods. The gray region represents constraints from microscopic theory~\cite{Dri16}.
}
\label{Ciampi}
\end{figure*}

Recently, new constraints~\cite{Ciampi+25} were obtained from the isospin transport ratio in $^{58,64}$Ni  + $^{58,64}$Ni collisions at 32 MeV/nucleon, where the impact parameter dependence was experimentally measured.
We understand that the
BUU transport model employs effective nuclear energy density
functionals that respect the present constraints from {\it ab initio} calculations. Figure~\ref{Ciampi} leaves little doubt on the ``outlier" nature of the PREX-II result.

\section{Conclusions and outlook}
\label{concl}

As noted in Ref.~\cite{S+A_2025}, the maximum-mass constraint moving to higher values, together with the causality requirement at any central density, poses significant restrictions on the high-density continuation of the EoS.  Based on those considerations, we selected a small number of EoS differing in the chiral order and/or the adiabatic index of their polytropic portion, followed by a parametrization of the pressure vs. density that guarantees causality. We also used a parametrization  where the speed of sound approaches the conformal limit at infinity and observed no significant impact on the M(R) relation or neutron star deformabilities.

We focused on the tidal deformability of neutron stars and related parameters, such as the tidal Love number. 
We addressed the effective tidal deformability of the binary and the implication of a very tight constraint on the chirp mass of the system.
Measurement of the effective tidal deformability (at nearly fixed mass) can be used to extract constraints on the neutron star radius. This is very valuable, as 
 precise neutron star radii would discriminate between equations of state better than any other single measurement. The upper limit on the effective tidal deformability from GW170817 excludes radii larger than 13.2 km, essentially independent of the assumed masses for the
component stars. This rules out the findings from PREX II and any steep EoS that generates large radii. We underline the importance of EoS-independent measurements, which can help interpret divergent outcomes from terrestrial experiments that rely on EoS modeling assumptions. Most recent constraints~\cite{Tang+25, Cui+25, Ciampi+25}, from both neutron star observations and heavy ion collisions at Fermi energy, confirm the soft nature of the nuclear equation of state.

One of our main takeaways is that, while phenomenological extensions of the EoS are unavoidable, they can and must be controlled. Extensions of the (soft) chiral EFT EoS are generally consistent with both terrestrial and observational constraints. With the new generation of GW detectors and continuing progress toward ab initio nuclear theory, there is real hope to build a robust bridge connecting microscopic physics and telescope physics. 

\section*{Acknowledgments}
This work was supported by 
the U.S. Department of Energy, Office of Science, Office of Basic Energy Sciences, under Award Number DE-FG02-03ER41270. \\

\section*{In Memoriam}
{\it F.S. dedicates this work to Rup, with everlasting love.}

\bibliography{bibRM}

\appendix
\section{Calculation of the gravitational Love number}

The spacetime of a non-rotating, spherically symmetric neutron star is described by the metric:
	\begin{equation}
		ds^2 = -e^{\nu(r)}dt^2 + e^{\lambda(r)}dr^2 + r^2(d\theta^2 + \sin^2\theta d\phi^2) \; ,
\label{der1}
	\end{equation}
where the metric function $\lambda(r)$ is related to the mass $m(r)$ enclosed within radius $r$ through
	\begin{equation}
		e^{\lambda(r)} = (1 - \frac{2m(r)}{r} )^{-1} \; ,
\label{der2}
	\end{equation}
and the hydrostatic equilibrium is governed by the TOV equations,
	\begin{equation}
\nonumber
		\frac{dm}{dr} = 4\pi r^2 \epsilon \; ,
\end{equation}
\begin{equation}
\nonumber
		\frac{d\nu}{dr} = \frac{2(m + 4\pi r^3 P)}{r(r - 2m)} \; ,
\end{equation}
\begin{equation}
		\frac{dP}{dr} = -\frac{1}{2}(\epsilon + P)\frac{d\nu}{dr} \; .
	\end{equation}

When a spherically symmetric star is placed in a quadrupolar tidal field $\epsilon_{ij}$, the response to the perturbation is a quadrupole moment, $Q_{ij}$, which, to linear order in $\epsilon_{ij}$, can be written as 
\begin{equation}
\label{QQ}
Q_{ij} = - \lambda \epsilon_{ij} \; .
\end{equation}
 Inserting the perturbed Einstein tensor~\cite{Koj1992} in the linearized Einstein equations,
one obtains a system of equations that reduces to a single second-order differential equation for a static, even parity $l=2$ metric perturbation $H(r)$,
\begin{equation}
 \label{hhh}
		H''(r) + Q_1(r)H'(r) + Q_0(r)H(r) = 0,
	\end{equation}
	with:
	\begin{equation}
		Q_1(r) = \frac{2}{r} + e^{\lambda} \big ( \frac{2m}{r^2} + 4\pi r (P - \epsilon)\big  )
\label{q1}
	\end{equation}
	and
	\begin{equation}
		Q_0(r) = e^{\lambda} \big ( -\frac{6}{r^2} + 4\pi (\epsilon + P) \frac{d\epsilon}{dP} + 4\pi (5\epsilon + 9P) \big ) - \big ( \frac{d\nu}{dr} \big )^2 \; .
\label{q0}
	\end{equation}
The solution of Eq.~(\ref{hhh}) is facilitated by the introduction of 
 the variable $y(r)$, defined as the dimensionless logarithmic derivative of $H(r)$:
	\begin{equation}
		y(r) = \frac{r H'(r)}{H(r)} \; .
\label{Ric}
	\end{equation}
Differentiating Eq.~(\ref{Ric}) with respect to $r$ and 
 substituting $\frac{H'}{H} = \frac{y}{r}$, we obtain
	\begin{equation}
		y' = \frac{y}{r} + r \frac{H''}{H} - \frac{y^2}{r} \; .
\label{yprim}
	\end{equation}
Solving Eq.~(\ref{hhh}) for the ratio $\frac{H''}{H}$ and substituting it in Eq.~(\ref{yprim}), we obtain:
	\begin{equation}
		y' = \frac{y}{r} + r \big (-Q_1 \frac{y}{r} - Q_0 \big ) - \frac{y^2}{r} 
		 = \frac{y}{r} - Q_1 y - r Q_0 - \frac{y^2}{r} \; ,
\end{equation}
and thus
\begin{equation}
		y' + \frac{1}{r} y^2 + y \big ( Q_1 - \frac{1}{r}\big ) + r Q_0 = 0 \; .
\label{y_prelim}
	\end{equation}
From Eq.~(\ref{q1}), we can write
	\begin{equation}
		Q_1 - \frac{1}{r} = \frac{2}{r} + e^{\lambda}  \big ( \frac{2m}{r^2} + 4\pi r (P - \epsilon) \big ) - \frac{1}{r} 
		 = \frac{1}{r} + e^{\lambda} \frac{2m}{r^2} + 4\pi r e^{\lambda} (P - \epsilon) \; ,
	\end{equation}
	which, together with the obvious identity
\begin{equation}
\frac{1}{r}= \frac{1}{r}e^{\lambda} \big  (1 - \frac{2m}{r} \big ) = \frac{e^{\lambda}}{r} - \frac{2m e^{\lambda}}{r^2} \; ,
\end{equation}
 yields
	\begin{equation}
		Q_1 - \frac{1}{r} 
		= \big  ( \frac{e^{\lambda}}{r} - \frac{2m e^{\lambda}}{r^2} \big )
		+ \frac{2m e^{\lambda}}{r^2} + 4\pi r e^{\lambda} \big  (P - \epsilon \big )
		= \frac{e^{\lambda}}{r} \big ( 1 + 4\pi r^2 (P - \epsilon) \big ) \; .
	\end{equation}
Substituting this back into Eq.~(\ref{y_prelim}) results in
	\begin{equation}
		y'(r) + \frac{1}{r}y^2(r) + \frac{1}{r}y(r)e^{\lambda(r)} \big [ 1 + 4\pi r^2 \big (P(r) - \epsilon(r) \big ) \big ] + rQ_0(r) = 0 \; ,
\label{yfin}
	\end{equation}
with
\begin{equation}
		Q_0(r) = e^{\lambda} ( -\frac{6}{r^2} + 4\pi (5\epsilon + 9P) + \frac{4\pi (\epsilon + P)}{c_s^2} ) - ( \frac{d\nu}{dr} )^2 \; ,
	\end{equation}
	and $ c_s^2 =  \frac{dP}{d\epsilon}. $ 

From the TOV equations, we have
	\begin{equation}
\label{tov}
		\big ( \frac{d\nu}{dr} \big )^2 =  \big( \frac{2(m + 4\pi r^3 P)}{r(r - 2m)} \big )^2 \; ,
	\end{equation}
	and, using $e^{\lambda(r)} = ( 1 - \frac{2m(r)}{r} )^{-1}$,
	\begin{equation}
\label{tov2}
		\big ( \frac{d\nu}{dr}  \big )^2 = \frac{4 e^{2\lambda}}{r^4} (m + 4\pi r^3 P)^2 \; .
	\end{equation}
Combining the terms, we arrive at
	\begin{equation}
\label{q0fin}
		Q_0(r) = 4\pi e^{\lambda} \big  ( 5\epsilon + 9P + \frac{\epsilon + P}{c_s^2}\big  ) - \frac{6e^{\lambda}}{r^2} - \frac{4 e^{2\lambda}}{r^4} [m + 4\pi r^3 P]^2 \; .
	\end{equation}

Finally, Eq.~(\ref{hhh}) for static, even-parity $l=2$ perturbations, reads:
	\begin{equation}
\label{hpp}
		H'' + H' \left[\frac{2}{r} + e^\lambda\left( \frac{2m}{r^2} + 4\pi\left(p-\epsilon\right)\right)\right] + H \left[4\pi e^{\lambda}\left(5\epsilon + 9p + \frac{\epsilon + p}{c_s^2}\right) - \frac{6 e^{\lambda}}{r^2} - \left(\frac{d\nu}{dr}\right)^2\right] = 0 \; .
	\end{equation}

	Next, we seek the solution at the center of the star, to which end we write Eq.~(\ref{hpp})
	in the limit of $r\rightarrow0$, 
	\begin{equation}
\label{inf}
		H'' + \frac{2}{r} H' - \frac{6}{r^2} H + 4\pi\left(5\epsilon(0) + 9 p(0) + \frac{\epsilon(0) + p(0)}{c_s^{2}(0)}\right) H = 0 \; ,
	\end{equation}
	and insert the trial solution
	\begin{equation}
\label{trial}
		H(r) = a_{0}r^{s} \; .
	\end{equation}
	 The regular solution at the center of the star has the form
	\begin{equation}
\label{reg}
		H(r) = a_{0}r^{2} + a_{2}r^{4} \; ,
	\end{equation}
	which we substitute in Eq.~(\ref{inf}) to obtain:
	\begin{equation}
		(2 a_0 + 12 a_2 r^2) + \frac{2}{r}(2 a_0 r + 4 a_2 r^3) - \frac{6}{r^2}(a_0 r^2 + a_2 r^4) + 4\pi\left(5\epsilon(0) + 9 p(0) + \frac{\epsilon(0) + p(0)}{c_s^{2}(0)}\right)(a_0 r^2) = 0 \; .
	\end{equation}
	After regrouping terms by powers of $r^2$, one finds a relation between $a_0$ and $a_2$ which we 
	substitute in Eq.~(\ref{reg}) to obtain
	\begin{equation}
		H(r) = a_0 r^2 \left[ 1 - \frac{4\pi}{14} \left( 5\epsilon(0) + 9p(0) + \frac{\epsilon(0)+p(0)}{c_s^2(0)} \right) r^2 \right] \; .
	\end{equation}
	
Next, we move to the vacuum exterior solution, r $>$ R. \\	
	For r $>$ R, a few straightforward steps lead to the second-order differential equation for $H(r)$,
	\begin{equation}
\label{hout}
		H'' + \frac{2\left(r-M\right)}{r\left(r-2M\right)} H' - \left[\frac{6}{r\left(r-2M\right)} - \left(\frac{2M}{r\left(r-2M\right)}\right)^{2}\right] H = 0 \; ,
	\end{equation}
	to which we apply the transformation $x = \frac{r}{M} -1$ .		
	The resulting equation,
\begin{equation}
\label{leg}
	\left(x^2-1\right) \frac{d^2 H}{d x^2}+2 x \frac{d H}{d x}-\left[6+\frac{4}{x^2-1}\right] H=0 \; ,
\end{equation}
is recognized as the associated Legendre equation for $l=m=2$, solved by a linear combination of the associated Legendre functions, $P_l^m(x)$ and $Q_l^m(x)$,
\begin{equation}
\label{legp}
 	H = c_1 Q_2^2(x) + c_2 P_2^2(x) \; .
\end{equation}
The exterior solution is then
\begin{equation}
\nonumber
	H= c_1\left[\frac{3}{2}\left(x^2-1\right) \ln \left(\frac{x+1}{x-1}\right)-\frac{3 x^3-5 x}{x^2-1}\right] +3 c_2\left(x^2-1\right) = 
\end{equation}
\begin{equation}
\label{ext}
 c_1\left(\frac{r}{M}\right)^2\left(1-\frac{2 M}{r}\right)\left[-\frac{M(M-r)\left(2 M^2+6 M r-3 r^2\right)}{r^2(2 M-r)^2}+\frac{3}{2} \ln \left(\frac{r}{r-2 M}\right)\right] 
  +3 c_2\left(\frac{r}{M}\right)^2\left(1-\frac{2 M}{r}\right) \; .
\end{equation}

To determine the coefficients $c_1$ and $c_2$, we need the asymptotic $H(r)$ solution. \\
For that purpose, we expand the various terms of the above equation as:
\begin{equation}
\label{term1}
	3 \left(\frac{r}{M}\right)^2\left(1-\frac{2 M}{r}\right)
	 =3\left(\frac{r}{M}\right)^2 +\mathcal{O}\left(\frac{r}{M}\right) \; ,
\end{equation}
\begin{equation}
\label{term2}
	\frac{3}{2}\ln \left(\frac{r}{r-2 M}\right)
	 =3 \frac{M}{r}+3 \frac{M^2}{r^2}+4 \frac{M^3}{r^3}+6\frac{M^4}{r^4}+\frac{48}{5} \frac{M^5}{r^5}+\cdots \; .
\end{equation}
Defining  $ \kappa = \frac{r}{M} $,
\begin{equation}
\label{term3}
\frac{M(M-r)\left(2 M^2+6 M r-3 r^2\right)}{r^2(2 M-r)^2} = \frac{(1-\kappa)\left(2+6 \kappa-3 \kappa^2\right)}{\kappa^2(2-\kappa)^2} \; ,
\end{equation} 
to which we apply the following expansion:
\begin{equation}
\label{kappa}
 \frac{(1-\kappa)\left(2+6 \kappa-3 \kappa^2\right)}{\kappa^2(2-\kappa)^2} =\frac{1}{\kappa}\left(3-\frac{9}{\kappa}+\frac{4}{\kappa^2}+\frac{2}{\kappa^3}\right)\left(1-\frac{4}{\kappa}+\frac{4}{\kappa^2}\right)^{-1} \; .
\end{equation} 
Expanding further the last factor in Eq.~(\ref{kappa}) as $\frac{1}{1 - x}$, with $x = \frac{4}{\kappa}-\frac{4}{\kappa^2}$, and after multiple algebraic manipulations, we arrive at
\begin{equation}
\frac{M(M-r)\left(2 M^2+6 M r-3 r^2\right)}{r^2(2 M-r)^2} = \frac{3}{\kappa}+\frac{3}{\kappa^2}+\frac{4}{\kappa^3}+\frac{6}{\kappa^4}+\frac{6}{\kappa^5}+\mathcal{O}\left(\kappa^{-6}\right) \; .
\label{exp}
\end{equation}
Combining all the expansions, the asymptotic solution for $H(r)$ becomes
\begin{equation}
\label{asy}
	H = \frac{8}{5} \left(\frac{M}{r}\right)^3 c_1 + \mathcal{O}\left(\left(\frac{M}{r}\right)^4\right) + 3\left(\frac{r}{M}\right)^2 c_2 + \mathcal{O}\left(\left(\frac{r}{M}\right)\right) \; .
\end{equation}
We are now in the position to determine the coefficients $c_1$ and $c_2$. In the star's local asymptotic rest frame, 
\begin{equation}
\label{gtt}
	\frac{1-g_{tt}}{2}=-\frac{M}{r}-\frac{3 Q_{i j}}{2r^3}\left(n^{i} n^{j} -\frac{1}{3} \delta^{ij}\right)+\frac{1}{2} \varepsilon_{ij} x^i x^j + ... \; ,
\end{equation}
where $n^i = x^i/r$, and therefore
\begin{equation}
	\frac{1}{2} \varepsilon_{ij} x^{i} x^{j} =\frac{1}{2} r^2 \varepsilon_{ij} n^{i} n^{j} \; .
\end{equation}
Using $\varepsilon_{i j}= \text{diagonal} \left(\frac{-\varepsilon}{2},\frac{-\varepsilon}{2},\varepsilon\right)$ and $n^i=(\sin \theta \cos \phi, \sin \theta \sin \phi, \cos \theta)$, 
\begin{equation}
\label{y20}
	\frac{1}{2} \varepsilon_{ij}x^{i} x^{j} =\frac{1}{2} r^2\left(\varepsilon_{xx} n_1^2+\varepsilon_{yy} n_2^2+\varepsilon_{zz} n_3^2\right) = 
\frac{\varepsilon}{2} r^2 \left[\frac{3\cos ^2 \theta-1}{2} \right] =
  \frac{1}{2} \varepsilon r^{2} Y_{20} \; .
\end{equation}
Similarly,
\begin{equation}
\label{pert}
\frac{3}{2} \frac{Q_{i j}}{r^3}\left(n^{i} n^{j}-\frac{1}{3} \delta^{ij}\right)= \frac{3Q}{2r^3} Y_{20} \; .
\end{equation}
Matching the asymptotic solution in Eq.~(\ref{asy}) to the asymptotic behavior of the metric component $g_{tt}$ in Eq.~(\ref{gtt}), and using $Q = -\lambda \varepsilon$, we find
\begin{equation}
\label{coef}
c_1 = \frac{15}{8}\frac{\lambda}{M^3} \varepsilon \; \; \; \; \; \; \; c_2 = \frac{1}{3}M^2 \varepsilon \, .
\end{equation}
Finally, using Eq.~(\ref{coef}) and Eq.~(\ref{ext}), we solve for $\lambda$ in terms of $H(r)$ and its derivative at $r=R$, which are then replaced in Eq.~(\ref{lam}) to yield the final expression:
\begin{equation}
\nonumber
k_2(C, y_R) = \frac{8}{5} C^5 (1 - 2 C)^2 [2 - y_R + 2C(y_R-1)] \{2C[6 -3y_R + 3C(5y_R - 8)] + 
\end{equation}
\begin{equation}
\nonumber
4C^3 [13 -11 y_R + C(3y_R - 2) + 2C^2(1 + y_R)] +
\end{equation}
\begin{equation}
3(1 - 2 C)^2[2 - y_R + 2C(y_R-1)] ln(1- 2 C)\}^{-1} \; ,
\end{equation}
where $ y = R \frac{H'(R)}{H(R)}$ is obtained integrating Eq.~(\ref{yfin}).

\end{document}